\documentclass[prd,aps,10pt,twocolumn,amsmath,superscriptaddress,showpacs,nofootinbib]{revtex4-1}

\usepackage[T1]{fontenc}
\usepackage[utf8]{inputenc}
\usepackage[brazil,english]{babel}
\usepackage{tensor}
\usepackage[usenames,dvipsnames]{xcolor}
  \definecolor{darkblue}{RGB}{0,0,150}
\usepackage{amssymb,amsthm}
\allowdisplaybreaks[1]
  \newtheorem{theorem}{Theorem}[section]

  \newtheorem{proposition}[theorem]{Proposition}
  
\usepackage{tensor}
\usepackage{graphicx}
\usepackage[ugly]{nicefrac}
\usepackage{siunitx}
\usepackage{hyperref}
\hypersetup{
  pdfinfo = {
    Author = {Daniel C. Guariento, Alan Maciel, Marina C. Mello, Vilson T. Zanchin},
    Title = {Charged cosmological black holes: a thorough study of a family of solutions}
  },
  colorlinks = true,
  breaklinks = true,
  linkcolor = darkblue,
  urlcolor = darkblue,
  citecolor = darkblue
}
\usepackage{paralist}

\usepackage[
            linecolor=orange,
            backgroundcolor=red!30,
            colorinlistoftodos]{todonotes}

\makeatletter
\providecommand*{\diff}%
{\@ifnextchar^{\DIfF}{\DIfF^{}}}
\def\DIfF^#1{%
  \mathop{\mathrm{\mathstrut d}}%
  \nolimits^{#1}\gobblespace}
\def\gobblespace{%
  \futurelet\diffarg\opspace}
\def\opspace{%
  \let\DiffSpace\!%
  \ifx\diffarg(%
  \let\DiffSpace\relax
  \else
  \ifx\diffarg[%
  \let\DiffSpace\relax
  \else
  \ifx\diffarg\{%
  \let\DiffSpace\relax
  \fi\fi\fi\DiffSpace}

\providecommand*{\deriv}[3][]{%
  \frac{\diff^{#1}#2}{\diff #3^{#1}}}

\DeclareMathOperator{\Lie}{\mathcal{L}}

\DeclareMathOperator{\diag}{diag}
\makeatother

\newcommand{\ud}{\ensuremath{\diff}}

\DeclareMathAlphabet{\mathpzc}{T1}{pzc}{m}{it}

\let\oldfrac\frac
\renewcommand{\frac}[2]{%
  \mathchoice
    {\oldfrac{#1}{#2}}
    {#1/#2}
    {#1/#2}
    {#1/#2}
}
\graphicspath{{figuras/}{../figuras/}}

\begin{document}

\title{Charged cosmological black holes: a thorough study of a family of solutions}

\author{Daniel C.\ Guariento}

\email{dguariento@perimeterinstitute.ca}

\affiliation{Perimeter Institute for Theoretical Physics, 31 Caroline St. N., Waterloo, Ontario N2L 2Y5, Canada} 

\affiliation{Conestoga College, 299 Doon Valley Drive, Kitchener, Ontario N2G 4M4, Canada}

\author{Alan Maciel}
\email{alan.silva@ufabc.edu.br}

\affiliation{Centro de Matemática, Computação e Cognição, Universidade Federal do ABC,\\
 Avenida dos Estados 5001, CEP 09210-580, Santo André, São Paulo, Brazil. }

\author{Marina M.\ C.\ Mello}

\affiliation{Centro de Ciências Naturais e Humanas, Universidade Federal do ABC,\\
 Avenida dos Estados 5001, CEP 09210-580, Santo André, São Paulo, Brazil. }

\author{Vilson T.\ Zanchin}

\affiliation{Centro de Ciências Naturais e Humanas, Universidade Federal do ABC,\\
 Avenida dos Estados 5001, CEP 09210-580, Santo André, São Paulo, Brazil. }

\begin{abstract}
  We study a class of charged cosmological black holes defined by the Shah--Vaidya solution, which is similar to the McVittie solution but for a central object of nonzero electric charge. We show that the Shah--Vaidya metric is a solution of Einstein's equations with a cuscuton and a Maxwell fields as sources, as well as a mass parameter. We then analyze the possible causal structures of the solution under some few physically reasonable assumptions, and determine the regions in the parameter space corresponding to well behaved charged cosmological black holes and those corresponding to naked singularities. The asymptotic behavior of the Hubble factor $H(t)$ is also determinant to the causal properties of the spacetime and a theorem explaining its effect is stated. Examples of causal diagrams covering all the possibles types of spacetimes allowed by our initial assumptions are drawn and discussed.
\end{abstract}


\maketitle 

\section{Introduction}

Black holes (BHs) have fascinated the general relativity (GR) community over the years. They represent solutions that, on the one hand, reach the theoretical applicability limit of GR  and, on the other hand, demand our most resourceful observational techniques in order to be detected \cite{Abbott:2016nmj, *Abbott:2016blz, Akiyama:2019eap,*Akiyama:2019brx, *Akiyama:2019sww, *Akiyama:2019bqs, *Akiyama:2019fyp}.

Stationary and quasi-stationary BH theory dates from the 1970s and uncovered a series of intriguing phenomena (BH thermodynamics, superradiance, Hawking radiation) \cite{carlip-2014,*bekenstein-1973a,*bekenstein-1998,*Brito:2015oca,*hawking-1975}. However, as far as we know, BHs in nature are formed by the collapse of matter under its own gravitational field. Therefore, they are dynamically generated objects that reach a stationary final state. In order to study the formation and evolution of BHs we have to consider solutions that evolve in time. A particularly tractable though still rich case of dynamical BH solutions is the black hole in an expanding universe, which has been the target of a significant number of studies \cite{thakurta1981,*Barrow:1991dn,*Sultana:2005tp,McClure:2006kg,Faraoni:2013aba}. 

The McVittie spacetime is one of the most representative exact solutions of such kind. It was first investigated in 1933 \cite{mcvittie-1932,*McVittie:1933zz} and describes a Schwarzschild black hole embedded in a Friedman--Lemaître--Robertson--Walker (FLRW) spacetime. It took many years after the solution was released in order for an appropriate assessment to be made about its causal structure. That delay may has happened because of the lack of mathematical tools available, since many important works on the machinery for the analysis of dynamical solutions were presented in the last two decades, namely Refs.\ \cite{Hayward:1993ph,*Hayward:1993mw,*Hayward:1994bu,*Hayward:1997jp,Senovilla:2011fk,*Mars:2003ud}. Since then, the McVittie solution has been extensively studied by many authors (see, e.g., Refs.\ \cite{Nolan:1998xs,*Nolan:1999kk,*Nolan:1999wf,*Nolan:2014maa,Carrera:2009ve,Kaloper:2010ec,Lake:2011ni,*anderson-2011,daSilva:2012nh,Landry:2012nv,Faraoni:2012gz}).  

Similarly to the Schwarzschild black hole, the McVittie spacetime is characterized by a mass parameter associated to a localized object. The addition of other parameters such as an electric charge and rotation has been considered in the literature. Of particular interest for the present work is the charged version of the McVittie metric given by Shah and Vaidya in 1968 \cite{shah-vaidya-1968}. Worth of mention at this point is also the extremely charged multiple cosmological black holes solution presented in Ref.\ \cite{kastor-traschen}, which is a generalization of the many charged black holes solution found by Hartle and Hawking \cite{Hartle:1972ya}. The causal structure of the solution presented in Ref.~\cite{kastor-traschen} was considered in Refs.\ \cite{Nakao:1994mm,Chimento:2013pka}. 

As well as the McVittie spacetime, charged cosmological black-hole spacetimes have gained some attention recently, as can be seen in Refs.\ \cite{Tang:2009hm,*Benone:2014qaa,Faraoni:2014nba,Rodrigues:2015sja,Bibi:2017urt}. The addition of charge can enrich the discussion by bringing new ingredients into play, which are useful in the study of more complex scenarios. For instance, it may have a qualitative behavior similar to asymptotically cosmological rotating black holes, since a non vanishing charge or angular momentum of the central object has similar effects on the spacetime geometry, as they introduce one more horizon behind the event horizon, in the non-extremal cases. 

In the present work, we investigate the charged McVittie spacetime, or Shah--Vaidya (SV) solution, by analyzing its fluid characteristics and its causal structure. Foremost we start by showing that an electromagnetic field plus a scalar field known as cuscuton \cite{Afshordi:2006ad,*Afshordi:2007yx} are sources of the SV solution. Afterward we analyze the causal structure, investigating the singularities and the horizons. We found that there are two main scenarios that may change with the parameters. Although the undercharged cases behave in a similar way to the neutral case in region outside the event horizon---we present a similar theorem relating the Hubble factor to the analytic continuation---in the overcharged case the causal structure is much different, with the presence of naked singularities. 
 

This paper is organized as follows: In Sec.\ \ref{sec:metric-iso}, we present the main elements of the Shah--Vaidya metric and its derivation as a solution of the Einstein--Maxwell equations for a central mass and electric charge in a self-gravitating cuscuton background. In Sec.\ \ref{sec:metric-areal}, we perform a change of variables to analyze the metric in areal radius coordinates, which is convenient in order to locate and characterize singularities and discontinuities throughout the various cases. In Sec.\ \ref{sec:causal}, we study the causal structure of the different cases and construct a few examples. We present our conclusions and next steps in Sec.\ \ref{sec:conclusion}.

Throughout this work, we employ reduced Planck units and the $(-,\,+,\,+,\,+)$ signature for the spacetime metric. Greek indices run from 0 to 3. We use isotropic spherical coordinates with the timelike coordinate denoted by $t$ and the spacelike radial coordinate denoted by $r$. When writing down differential equations, the primes represent partial derivatives with respect to the $r$ coordinate, and overhead dots represent partial derivatives with respect to the $t$ coordinate.

\section{A charged cosmological black hole}\label{sec:metric-iso}

\subsection{The metric in isotropic coordinates}

The Shah--Vaidya line element, in isotropic coordinates, is given by \cite{shah-vaidya-1968,sussman-1987,*sussman-1988a,*sussman-1988b,Gao:2004cr,McClure:2006kg,Faraoni:2014nba,Rodrigues:2015sja}
\begin{equation} \label{eq:metric1}
  \begin{split}
    \diff s^2 =& - \frac{\left[1 -\mu^2 +\chi^2\right]^2}{\left[ \left( 1 + \mu \right)^2 - \chi^2 \right]^2} \diff t^2 \\
    & + a^2(t) \left[ \left(1 + \mu\right)^2 - \chi^2 \right]^2 \left( \diff r^2 + r^2 \diff \Omega^2 \right) \,,
  \end{split}
\end{equation}
where $a(t)$ is the scale factor and the functions $\mu$ and $\chi$, are defined respectively as
\begin{subequations}\label{eq:auxiliary}
  \begin{align}
    \mu =&\, \mu (t, r)
    \equiv \, \frac{m}{2r a(t)} \,,\\
    \chi =&\, \chi (t, r)
    \equiv\, \frac{q}{2 r a(t)} \,.
  \end{align}
\end{subequations}
The parameters $m$ and $q$ are to be associated to the mass and electric charge of a localized source. 

The presence of a pointlike electric charge gives rise to an electromagnetic Faraday--Maxwell field strength which may be written as   
%
\begin{equation} \label{eq:FMfield1}
  F_{\mu \nu} = \frac{q}{r^2 a(t)}\frac{\left[1 - \mu^2 + \chi^2 \right]} {\left[ \left(1+ \mu\right)^2 - \chi^2 \right]^2 } \left( \delta^t_\mu \delta^r_\nu - \delta^r_\mu \delta^t_\nu \right),
\end{equation}
where $\delta _\mu^\nu$ is the Kronecker delta tensor.

\subsection{Field sources}

In this section, we show that the line element \eqref{eq:metric1} is a solution to the coupled Einstein--Maxwell equations with the homogeneous cuscuton field as a source and a central object with electric charge. The procedure is analogous to the one developed in Refs.\ \cite{shah-vaidya-1968,Abdalla:2013ara}. 

The cuscuton field has been proposed as non-canonical field theory that has no dynamical degree of freedom, but that is able to change the dynamics of other fields, including gravity, when coupled to them \cite{Afshordi:2006ad,*Afshordi:2007yx,Gomes:2017tzd}. The action for the cuscuton scalar $\phi$ is given by
\begin{equation}
  S_\phi = \int \diff^4 x \sqrt{-g} \left[ \mu^2 \sqrt{\left| 2 X \right|} - V (\phi) \right],
\end{equation}
with $V (\phi)$ being the scalar potential, $\mu$ the cuscuton coupling, and $X$ the usual scalar kinetic term given by 
\begin{equation}
  X \equiv -\frac{1}{2} g^{\mu \nu} \nabla_\mu \phi \nabla_\nu \phi.
\end{equation}
The second source is a nonzero electromagnetic field, whose action is given by
\begin{equation}
  S_\text{EM} = \int \sqrt{-g} \diff^4 x \left[ -\frac{1}{16\pi} F_{\mu \nu} F^{\mu \nu} + A_\mu J^{\mu} \right],
\end{equation}
so the full action is 
\begin{equation}\label{eq:action-full}
  S = \dfrac{1}{16\pi}\int \sqrt{-g}\diff^4 x \, \mathcal{R} + S_\text{EM} + 
S_\phi.
\end{equation}

We assume a spherically symmetric ansatz for the metric, with the following generic line element:
\begin{equation}
  \diff s^2 = -e^{2 \nu (t,r)} \diff t + e^{2 \lambda (t,r)} \diff r^2 + Y^2 (t,r) \diff \Omega^2. \label{eq:gen-sph-sym}
\end{equation}
In such a metric, the free electromagnetic field quantities $ F_{\mu \nu}$ reduce to a single function, denoted by $E(t,r)$, where
\begin{equation}
  F_{\mu \nu} = E (t,r) \left( \delta^t_\mu \delta^r_\nu - \delta^r_\mu \delta^t_\nu \right).
\end{equation}
We assume a comoving current in this frame, that is,
\begin{equation}
  J^{\mu} = j (t,r) u^{\mu} \,,
\end{equation}
where $u^\mu \equiv e^{-\nu} \delta_t^\mu$ is the comoving flow in metric \eqref{eq:gen-sph-sym}. Since the cuscuton field is assumed to be neutral and there are no other couplings, the Maxwell equations read
\begin{subequations}
  \begin{align}
    \dot{E} + E \left( 2 \frac{\dot{Y}}{Y} - \dot{\lambda} - \dot{\nu} \right) = &\,0, \label{eq:mer}\\ 
    E' + E \left( 2 \frac{Y'}{Y} - \lambda' - \nu' \right) =&\, -4\pi\,j e^{\nu + 2\lambda}. \label{eq:met}
  \end{align}
\end{subequations}
Eq.\ \eqref{eq:mer} can be integrated to give
\begin{equation}\label{EM-sol}
  E = q \frac{e^{(\nu + \lambda)}}{Y^2} \,,
\end{equation}
where $q = q (r)$ is an integration factor that may depend on the radial coordinate $r$. Inserting this solution for $E(t,r)$ into Eq.\ \eqref{eq:met}, we find
\begin{equation}
  q' = -4\pi\, j \, Y^2 e^{\lambda}.
\end{equation}
By assuming that there is no charge distribution in the bulk, we set $j = 0$ and then integrate the last equation to find that, in such a case, $q$ is also a constant throughout $r$.  

The equation of motion for the homogeneous cuscuton field $\phi = \phi (t)$ is cast from the action \eqref{eq:action-full} with metric \eqref{eq:gen-sph-sym} as \cite{Abdalla:2013ara} 
\begin{equation}\label{cusc-solution}
  \left( \frac{2 \dot{Y}}{Y} + \dot{\lambda} \right) e^{-\nu} = - \frac{1}{\mu^2} \deriv{V}{\phi} \equiv\, 3 H (t) \,.
\end{equation}
As in the uncharged case, this result means that the foliation defined by the flow $u^\mu$ (which also coincides with the direction defined by the field) has a constant mean extrinsic curvature (CMC), that is,
\begin{equation}\label{eq:CMC}
    \tensor{K}{^\mu _\mu} \equiv \frac{1}{2} g^{\mu \nu} \Lie_u \left( g_{\mu \nu} + u_\mu u_\nu \right) = 3 H(t),
\end{equation}
where $\Lie_u$ stands for the Lie derivative, and $H(t)$ is the same homogeneous function as defined in Eq.\ \eqref{cusc-solution}, which also tells us that the expansion scalar associated with the flow $u^\mu$ is independent of the radial coordinate.

Finally, the Einstein equations read
\begin{widetext} 
  \begin{subequations}
    \begin{align}
 &    \dot{Y}' - Y' \dot{\lambda} -  \dot{Y} \nu' = 0 ,\label{EEtr}\\
 &     \left[ \frac{2 \dot{Y} \dot{\lambda}}{Y} + \left( \frac{\dot{Y}}{Y} 
\right)^2 \right] e^{-2 \nu} + \left[ \frac{2 \left( \lambda' Y' - Y'' 
\right)}{Y} - \left( \frac{Y'}{Y} \right)^2 \right] e^{-2 \lambda} + 
\frac{1}{Y^2} =   \frac{q^2}{Y^4} + V, \label{EEtt}\\
  &    \left[ \frac{2 \left(\ddot{Y} - \dot{\nu} \dot{Y}  \right)}{Y} + \left( 
\frac{\dot{Y}}{Y} \right)^2 \right] e^{-2 \nu} - \left[ \frac{2 \nu' Y'}{Y} +
\left( \frac{Y'}{Y} \right)^2 \right] e^{-2 \lambda} + \frac{1}{Y^2} =
 \frac{q^2}{Y^4} - \mu^2 e^{-\nu} \dot{\phi} + V, \label{EErr}\\
 &     \left[ \dot{\nu} \dot{\lambda} - \ddot{\lambda} - \dot{\lambda}^2 + 
\frac{\dot{Y} \left( \dot{\nu} - \dot{\lambda} \right) - \ddot{Y}}{Y} \right] 
e^{-2 \nu} + \left[ \nu'' + \nu^{\prime 2} - \nu' \lambda' + \frac{ Y' \left( 
\nu' - \lambda' \right) + Y''}{Y} \right] e^{-2 \lambda} = \frac{q^2}{Y^4} + 
\mu^2 e^{-\nu} \dot{\phi} - V \label{EEthth}.
    \end{align}
  \end{subequations}
\end{widetext}

Further imposing that the traceless part of the extrinsic curvature vanishes, that is, that the comoving flow be shear-free\footnote{Imposing shear-free flow in a spherically symmetric metric implies a CMC comoving foliation, but the converse is not true.}, we find the relations
\begin{align}
  \dot{\lambda} =&\, \frac{\dot{Y}}{Y},\\
  Y =&\, \eta (r) e^{\lambda},
\end{align}
where $\eta (r)$ is a function that depends only on the radial coordinate. Inserting these conditions and Eq.\ \eqref{eq:CMC} into the momentum constraint \eqref{EEtr}, we find 
\begin{equation}\label{enu}
  e^{\nu} = \frac{\dot{\lambda}}{H}.
\end{equation}
With this, the metric \eqref{eq:gen-sph-sym} can be cast in the form
\begin{equation}\label{eq:metric-shearfree}
  \diff s^2 = -\left( \frac{\dot{\lambda}}{H} \right)^2 \diff t^2 + e^{2 \lambda} \left[ \diff r^2 + \eta^2(r) \diff \Omega^2 \right].
\end{equation}

An important difference with respect to the uncharged Kustaanheimo--Qvist class of shear-free, perfect-fluid metrics is that here, because of the presence of the electromagnetic field, the pressure isotropy condition is no longer satisfied. This means that, in the present case, the difference between the radial and angular components of the Einstein equations is not zero. Instead, after subtracting Eq.\ \eqref{EErr} from \eqref{EEthth}, we now find an anisotropy source term that reads
\begin{multline}\label{isotropy}
  -\lambda^{\prime 2} - 2 \frac{\lambda' \dot{\lambda}'}{\dot{\lambda}} + \lambda'' + \frac{\dot{\lambda}''}{\dot{\lambda}} - \frac{\eta'}{\eta} \left( \lambda' + \frac{\dot{\lambda}'}{\dot{\lambda}} \right)\\
  + \frac{1}{\eta^2} \left( \eta\,\eta'' - \eta^{\prime 2} + 1 \right) = - \frac{2 q^2 e^{-2 \lambda}}{\eta^4}\,,
\end{multline}
so that we are motivated to choose a gauge for which \cite{sussman-1987,mcvittie-1984}
\begin{equation}\label{Rdef}
  \eta \, \eta'' - \eta^{\prime 2} + 1 = 0 \quad \Rightarrow \quad \eta^{\prime 2} =   1 - k \eta^2 ,
\end{equation}
where $k$ is an integration constant that may be normalized to $k =0, \, \pm 1$. Integrating this equation, we get
\begin{equation}\label{Rcases}
  \eta (r) =
  \begin{cases}
    \sinh r & k = -1, \\
    r & k = 0, \\
    \sin r & k = 1.
  \end{cases}
\end{equation}

Equation \eqref{isotropy} multiplied by the factor $\dot{\lambda} e^{\lambda}$ can be rewritten as a total derivative, namely, 
\begin{equation}
  \partial_t \left[ e^{\lambda} \left( \lambda'' - \lambda^{\prime 2} - \lambda'\frac{\eta'}{\eta} \right) + \frac{2 q^2 e^{-2 \lambda}}{\eta^4} \right] = 0,
\end{equation}
which reduces to the ordinary differential equation:
\begin{equation}\label{isotropy_R}
  e^{\lambda} \left[ \lambda'' - \lambda^{\prime 2} -\lambda' \frac{\eta'}{\eta} + \frac{2 q^2 e^{-2 \lambda}}{\eta^4} \right] = \psi (r),
\end{equation}
where $\psi (r)$ is an arbitrary function of the radial coordinate alone.

With these results, we note that the Weyl part of the Misner--Sharp mass now reads \cite{carrera-rmp-2010,Faraoni:2015cnd,Guariento:2016hqo} 
\begin{equation}\label{eq:misnersharp}
  M_{\text{MS}}^{\text{(W)}} = \frac{\eta^3(r)}{3} \psi(r) - \frac{e^{-\lambda(t,r)}}{\eta(r)} q^2.
\end{equation}
Since we are interested in cosmological solutions, we choose the function $\eta(r)$ that provides a spatially flat metric, that is, $k = 0$, and, by Eq.\ \eqref{Rcases}, $\eta = r$. Now, the second term of Eq.\ \eqref{eq:misnersharp} is due to the presence of the central charge $q$, whereas the first term is related to the presence of a central mass. For that reason, we make the following ansatz for $\psi$  \cite{Guariento:2016hqo}, which holds for a spatially flat asymptotic FLRW spacetime\footnote{A more generic ansatz is $\psi = 3 m \frac{w^{\prime 2} (r)}{w^3 (r)}$, with $w (r) = 2 \eta (\frac{r}{2})$, which would recover the properties of cosmological spacetimes of any spatial curvature, as shown in Ref.\ \cite{Guariento:2016hqo} by assuming that the source is an otherwise unconstrained perfect fluid. However, as it was shown in Ref.\ \cite{Abdalla:2013ara}, a metric of the family \eqref{eq:metric-shearfree} with $k \neq 0$ is not a solution of the cuscuton equation of motion, Eq.\ \eqref{cusc-solution}.},
\begin{equation}
  \psi(r) = \frac{3 m}{r^3}.
\end{equation}
With such a choice, it can be shown that the following expression is a solution to Eq.\ \eqref{isotropy_R}:
\begin{equation}\label{elambda}
  e^\lambda = a (t) \left[ \left( 1 + \frac{m}{2 a (t) r} \right)^2 -  \left( \frac{q}{2 a (t) r} \right)^2 \right],
\end{equation}
where $a (t)$ is an arbitrary function of the coordinate $t$ alone. Considering the definition of the trace of the extrinsic curvature, Eq.\ \eqref{eq:CMC}, and bearing in mind that we are looking for cosmological solutions, we identify
\begin{equation}\label{eq:hubble}
  H = \frac{\dot{a}}{a}.
\end{equation}
This identification and the metric function~\eqref{elambda} substituted into Eq.~\eqref{enu} furnish the metric coefficient $e^\nu$,
\begin{equation}\label{enu2}
  e^\nu =  \frac{  1 - \left[\frac{m}{2 a (t) r} \right]^2 +  \left[ \frac{q}{2 a (t) r} \right]^2} { \left[ 1 + \frac{m}{2 a (t) r} \right]^2 -  \left[ \frac{q}{2 a (t) r} \right]^2 }.
\end{equation}
Applying this solution to the electromagnetic field \eqref{EM-sol}, we find
\begin{equation}
  E = \frac{q}{a r^2} \frac{1 - \frac{m^2}{4 a^2 r^2} + \frac{q^2}{4 a^2 r^2} }{\left[ \left( 1 + \frac{m}{2ar} \right)^2 - \frac{q^2}{4a^2 r^2} \right]^2},
\end{equation}
which is consistent with Eq.\ \eqref{eq:FMfield1}.

The remaining independent Einstein equations, \eqref{EEtt} and \eqref{EErr}, finally read 
\begin{align}
  & V - 3 H^2= 0,\\
 &  V - 3 H^2 = \left( \mu^2 \dot{\phi} + 2\dot{H} \right) \frac{\left( 1 +  \frac{m}{2ar}\right)^2 - \frac{q^2}{4a^2 r^2}}{1- \frac{m^2 - q^2}{4a^2 r^2}},
\end{align}
so that the cuscuton field satisfies the relation $\mu^2 \dot{\phi} + 2\dot{H} = 0$. Combining this relation with the equation of motion of the cuscuton field, Eq.\ \eqref{cusc-solution}, we find the same solution for the potential as in \cite{Abdalla:2013ara}, \emph{i.e.},
\begin{align}
  \left( \deriv{V}{\phi} \right)^2 =&\, 3 \mu^4 V \,,\\
  V =&\, \frac{3 \mu^4}{4} \left( \phi - V_0 \right)^2 \,, 
\end{align}
and it is then seen that the entire system is consistent.

In summary, we have shown that the Shah--Vaidya metric is an exact solution of the Einstein--Maxwell equations with a neutral cuscuton field source with a quadratic potential, and that it represents a central object with Misner--Sharp mass $m$ and electric charge $q$ in a spatially flat environment.

\section{Properties of the Shah--Vaidya spacetime}\label{sec:metric-areal}

\subsection{The areal radius coordinate}

The Shah--Vaidya metric and related physical and geometric quantities assume simpler forms when expressed in terms of the areal radius $R$, defined from metric \eqref{eq:metric1} as 
\begin{equation} \label{eq:arealR}
  \begin{split}
    R =&\, a r \left[ (1 + \mu)^2 - \chi^2 \right] 
    =\, a r + m + \frac{m^2 - q^2}{4 a r},
  \end{split}
\end{equation}
which transforms the line element \eqref{eq:metric1} into the form
\begin{equation}\label{eq:metric-areal}
  \ud s^2 = - N^2 \diff t^2 + \left[ \frac{\ud R}{N} - H R \diff t \right]^2 + R^2 \diff \Omega^2 ,
\end{equation}
where $N = N (R)$ is the lapse function, given by
\begin{equation}\label{eq:lapse}
  N = \sqrt{1 -\dfrac{2m}{R} + \dfrac{q^2}{R^2}}.
\end{equation}

Notice that the transformation~\eqref{eq:arealR} is defined in the real domain for $r$ and $R$ if $1 - \frac{2m}{R} + \frac{q^2}{R^2} \geq 0$. In fact, solving Eq.~\eqref{eq:arealR} for $r$, one gets two branches, namely \cite{Faraoni:2014nba}, 
\begin{equation}
  2 a \, r = R - m \pm R\sqrt{1 - \frac{2m}{R} + \frac{q^2}{R^2}},
  \label{eq:isotropicR}
\end{equation}
where the positive branch is the one that corresponds to $R \to \infty$ as $r \to \infty$ and the interpretation of the negative branch is given in the following sections. 

As just mentioned, the isotropic radius $r$ is real only if $1 - \frac{2 m}{R} + \frac{q^2}{R^2} > 0$. Even though this is well known for the relation between isotropic and areal radial coordinates in the Reissner--Nordström metric, we review some details here for completeness and for future reference. There are three different cases to be considered: \begin{inparaenum}[(i)] \item the undercharged, \item the extremely charged, and \item the overcharged cases. \end{inparaenum}

\subsubsection{Undercharged case: $m^2 > q^2$}

In this case, the lapse function $N (R)$ has two roots $R_\pm = m \pm \sqrt{m^2 - q^2}$, and $N^2 (R)$ is negative in the interval $R_-< R < R_+$. Hence, considering that the isotropic coordinate is real, the domain of the areal radius coordinate $R$ should be restricted to the intervals $R_+ \leq R < \infty$ and $0 \leq R \leq R_-$. If one assumes, as usual, that the areal radius ranges from zero to infinity, $R\in [0,\,\infty)$, then we conclude that, for $m^2 > q^2$, the isotropic coordinate $r$ does not cover the interval $R_-< R < R_+$. The region $R\in \left[ R_+, \, \infty \right)$ is covered once by $a r \in \left[ \frac{\sqrt{m^2 - q^2}}{2}, \, \infty \right)$  and again (twice) by $ a r \in \left[ \frac{\sqrt{m^2 -q^2}}{2},\, 0 \right)$. On the other hand, the region $R \in \left[ 0, \, R_- \right]$ is covered once by $a r \in \left[ - \frac{\left( m + |q| \right)}{2}, -\frac{\sqrt{m^2 - q^2}}{2} \right)$ and again (twice) by $ a r \in \left[ \frac{(-m + |q|)}{2}, \, -\frac{\sqrt{m^2 - q^2}}{2} \right)$.

\subsubsection{Extremal case: $m^2 = q^2$}

In this case, $N(R)$ has a double root $R_+ = R_- = m = |q|$, and $N^2 (R)$ is non-negative in the whole domain $0 \leq R < \infty$, where it is assumed, as usual, that the areal radius ranges from zero to infinity. We then conclude that, for $m^2 = q^2$, the isotropic coordinate $r$ covers (once) all the range of $R$. Equation \eqref{eq:arealR} reduces to $R = a r + m$, and so the region $R \in \left[ R_+ = R_-, \, \infty \right)$ is covered once by $a r \in \left[ 0, \, \infty \right)$. On the other hand, the region $R \in \left[ 0, \, R_- = R_+ \right]$ is covered once by $a r \in \left[ -m, 0 \right)$.

\subsubsection{Overcharged case: $m^2 < q^2$}

In this case, $N (R)$ has no real roots and $N^2 (R)$ remains nonzero in the whole domain $0 \leq R < \infty$, where it is assumed, as usual, that the areal radius ranges from zero to infinity. Therefore, for $m^2 < q^2$, the isotropic coordinate $r$ covers (twice) all the range of $R$. Indeed, the full range $R \in \left[ 0, \, \infty \right)$ is covered once by $a r \in \left[ \frac{(-m + |q|)}{2}, \, \infty \right)$, and again (twice) by $a r \in \left[ \frac{(-m 
-|q|)}{2}, 0 \right)$.

\subsection{The scale factor}

Here we assume that the function $a (t)$, which plays the role of the scale factor, implies a big-bang-type expanding model, and consider a few different asymptotic behaviors at late times. The important function in the present analysis is the Hubble factor as defined in Eq.\ \eqref{eq:hubble}. In view of this fact, we enumerate here the assumptions on $H (t)$.

\begin{enumerate}[(i)] 
\item Big-bang hypothesis: $\displaystyle{\lim_{t\to 0} H(t) \to + \infty}$. 
\item Null energy condition satisfied: $\dot H(t) < 0$.
\item Expanding hypothesis: $H (t) \geq 0$: 
  \begin{enumerate}
  \item Asymptotically de~Sitter cosmological model: $H (t) \to H_0 =$ constant at large $t$.
  \item Asymptotically empty model: $H(t) \to 0 $ at large $t$. See Appendix \ref{appendixH0zero}.
  \end{enumerate}
\end{enumerate}

\subsection{Curvature scalars}
\label{sec:scalars}

In order to understand the structure of the Shah--Vaidya spacetime, we have to describe the curvature singularities displayed by the metric~\eqref{eq:metric-areal}. Part of this analysis can be found in previous works, \emph{e.g.}, \cite{McClure:2006kg,Faraoni:2014nba,Rodrigues:2015sja}. For such a purpose, we make use of scalar invariants such as the Ricci $\mathcal{R}$ and the Kretschmann $\mathcal{K}$ scalars. These may be written respectively in the form 
\begin{align}
  \mathcal{R} =&\, 12 H^2 (t)+ \frac{6 \dot{H}}{N (R)}, \label{eq:ricci}\\
  \begin{split}\label{eq:kretsch}
    \mathcal{K} =&\, 48 \left( \frac{m}{R^3} - \frac{q^2}{R^4} \right)^2 + \frac{8q^4}{R^8} + 24 H^4 \\
    & + \frac{4 \dot{H}}{N (R)} \left( \frac{3 \dot{H}}{N} + {6 H^2} - \frac{2 q^2}{R^4} \right),
  \end{split}
\end{align}
where $N(R)$ is the lapse defined in Eq.\ \eqref{eq:lapse}. 

Assuming that the areal radius is restricted to the interval $0 \leq R < \infty$, it is clearly seen that singularities may occur at $R = 0$, or at regions of the spacetime where $N (R) = 0$, besides the singularities for which $H$ and/or $\dot{H}$ become arbitrarily large. Such a situation happens for instance if the scale factor is assumed to have a big-bang-like behavior. Therefore, the important function to investigate here is $N (R)$, whose real positive roots lead to curvature singularities. On the other hand, if $N (R)$ does not vanish anywhere, the singularity lies at $R = 0$.  

Let us then assume that $N (R)$ vanishes  and let us call $\mathcal{S}_0$ the corresponding locus, which corresponds in general to a spherical surface (of constant $R$) in the spacetime. The loci where $N (R)$ vanishes are given by the two spherical surfaces, $\mathcal{S}_\pm$, defined by $R = R_\pm$, where 
\begin{equation}
  R_\pm = m \pm \sqrt{m^2- q^2}, \label{re:singularities}
\end{equation}
which are real for $m^2\geq q^2$.

It is straightforward to show that the singularities $\mathcal{S}_\pm$ are both spacelike surfaces if the roots of $N(R)$ are real. Indeed, taking $n_\mu$ as the gradient of the surface ${\cal S} = R = \text{constant}$, namely, $n_\mu = \nabla_\mu \mathcal{S}$, we  have that  
\begin{equation}\label{eq:normS}
  n^2 = n^\mu n_\mu = N^2 - H^2R^2.
\end{equation}
Taking now the limit $R \to R_\pm$, it follows that $n^2 = n_\pm^2 = - H^2 R_\pm^2 < 0 $. Hence, the normal vectors $n^\mu_\pm $ are both timelike vectors and consequently $\mathcal{S}_\pm$ are both spacelike surfaces.

Analogously, it can be shown that the singularity at $R = 0$ is timelike. Indeed, following the same procedure as above, we may take the limit of small $R$ into Eq.\ \eqref{eq:normS} to obtain $n^2 \to q^2/R^2 >0$, showing that that $n^\mu $ is a spacelike vector so that $R=0$ is a timelike surface.

To proceed further, it is convenient to split the study of singular surfaces in the following cases.

\subsubsection{Undercharged case: $m^2 > q^2$ }\label{sec:sing-m<q}

Let us choose the positive branch of the isotropic radial coordinate [see Eqs.\ \eqref{eq:arealR} and \eqref{eq:isotropicR}] for which the $a (t) \, r \to 0$ limit corresponds to $R \to \infty$, while $R = 0$ corresponds to $a (t) \, r = \frac{\left( -m \pm |q| \right)}{2}$, where the isotropic coordinate $r$ is allowed to assume negative values. Is this case, when $m^2 > q^2$, the function $N (R)$ has two real positive roots for $R$, $R_\pm = m \pm \sqrt{m^2-q^2}$, both corresponding to singularities of the curvature scalars, besides the central singularity at $R = 0$. 

The presence of curvature singularities requires that the initial range $0 \leq R < \infty$ be reconsidered. In fact, the original range may be split into three intervals, separated by points (surfaces) where the curvature scalars diverge, namely,
\begin{inparaenum}[(i)]
\item\label{R+inf} $R_+ < R < \infty$,
\item\label{R-R+} $R_- < R < R_+$, and
\item\label{0R-} $0 < R < R_-$.
\end{inparaenum}
Since the metric may describe a cosmological spacetime, the asymptotic region $R \to \infty$ is of interest. So we will restrict our analysis to the interval $R_+ < R < \infty$, with  $R_+$ representing a boundary of the spacetime. The regions $R = R_-$ and $R = 0$ are also curvature singularities, but are of little interest for the present analysis, since they are beyond the singular spacetime boundary at $R = R_+$. Of course, the coordinate patches (\ref{R-R+}) and (\ref{0R-}) in the intervals $R_- < R < R_+$ and $0 < R < R_-$, respectively, correspond to disconnected spacetimes whose boundaries are singular surfaces in the sense that the curvature scalars diverge at those boundaries. None of these spacetime patches are considered in the present work.

As shown above, the spacetime boundary $R \to R_+$ is a spacelike singular surface, and we notice here that, apart from the roots of the scale factor $a (t)$, it is the only curvature singularity. The causal structure of these spacetimes is studied in Sec.\ \ref{sec:causal}.

\subsubsection{Extremely charged case $m^2 = q^2$ }\label{sec:sing-m=q}

This is a particular but very interesting situation for these solutions. First of all, Eq.\ \eqref{eq:arealR} becomes $R = a (t) \, r + m$, so that $R = 0$ corresponds to $r = \frac{-m}{a}$. The lapse function $N (R)$ has only one positive root, $R = R_\pm = m$, which corresponds to a singularity of the curvature scalars. The relevant spacetimes are obtained by choosing $R$ in the range $m < R < \infty$. This branch presents a singular boundary at $R \to m$. The causal structure of these spacetimes are studied in Sec.\ 
\ref{sec:causal}. 

The other region, namely, $0 < R < m$, represents another spacetime with singular boundaries at $R = 0$ and $R = m$, which we do not consider here.

\subsubsection{Overcharged case: $m^2< q^2$}

In this case, the lapse function has no roots and there is a curvature singularity at $R = 0$. The spacetime described by the overcharged SV metric includes all the range of the radial coordinate $0 < R < \infty$ and is bounded by a singular timelike surface at $R = 0$. 

Similarly to the static overcharged Reissner--Nordström--de~Sitter (RNdS) case, depending on the asymptotic form of the scale factor $a (t)$, there may be interesting causal structures here. This is also investigated in Sec.\ \ref{sec:causal}.

\section{Causal Structure}\label{sec:causal}

The aim of this section is to characterize the family of spacetimes represented by the Shah--Vaidya metric and determining their causal structure, which depends on some specified parameters and on the asymptotic behavior as it will be seen next. Our approach is to determine the loci of interest:

\begin{itemize}
\item Curvature singularities;
\item Apparent horizons (AHs);
\item End points of incomplete geodesics. 
\end{itemize}

Since the singularities were already approached in Sec.~\ref{sec:metric-areal}, in this section we study the AHs and geodesic completeness, in order to determine when an analytical continuation is possible. Finally, we characterize those continuations by means of a theorem and display a few representative examples.

\subsection{Existence and number of apparent horizons}

The apparent or trapping horizons are defined here following the approach proposed by Ref.\ \cite{Hayward:1993mw}, based on a study of the optical scalars related to null congruences defined by a codimension-two foliation of the spacetime. This so-called dual null approach has been used in a variety of applications in black hole physics \cite{Hayward:1994bu,Hayward:1997jp,Mars:2003ud,Senovilla:2011fk,Dafermos:2004wr,schnetter-2006} and beyond \cite{Faraoni:2013aba,Maciel:2015vva,Maciel:2015ypv,Faraoni:2015sja,Maciel:2018tnc}. 

We define AHs as the loci where one of the null expansions related to ingoing or outgoing null geodesics vanish. In spherically symmetric spacetimes like the Shah--Vaidya case, the AHs are 3-dimensional hypersurfaces that evolve in time, corresponding to a sphere at each time, and the analysis of the null expansion is reduced to the analysis of $\deriv{R_\pm(t)}{t}$, where $R_\pm (t)$ are outgoing/ingoing null geodesics. This allows us to classify each spacetime sphere as being

\begin{enumerate}[(a)]
 \item regular, if $(\deriv{R_+}{t})\,( \deriv{R_-}{t}) < 0$;
 \item trapped, if $(\deriv{R_+}{t})\,( \deriv{R_-}{t})  > 0$ and both are 
negative;
 \item anti-trapped, if $(\deriv{R_+}{t})\,( \deriv{R_-}{t})  > 0$ and both are 
positive;
\item marginal, if $(\deriv{R_+}{t})\,( \deriv{R_-}{t})  = 0$. 
\end{enumerate}

From the above definitions, it follows that the apparent horizons (AHs) are loci defined by the continuous sets of marginal spheres. From this, and assuming continuity of all quantities, we also see that the AHs are the boundaries between different regions.

The equations for outgoing/ingoing null geodesics are, respectively,
\begin{equation}\label{eq:null-geod-pm}
  \deriv{R_\pm}{t} =\, N (R_\pm) \left( R_\pm H (t) \pm N (R_\pm) \right),
\end{equation}
where we can readily notice that only $\deriv{R_-}{t}$ may vanish. Therefore, all the apparent horizons (AHs) of the Shah--Vaidya metric \eqref{eq:metric-areal} are given by the real and positive solutions of the equation 
\begin{equation} \label{eq:AH1}
  \begin{split}
    N (R) - H(t) R = 0 \Leftrightarrow N(R) ^2 - H(t)^2 R^2 =&\, 0 \Leftrightarrow \\
    -H^2 (t) R^4 + R^2 - 2 m R + q^2 =&\, 0.
  \end{split}
\end{equation}
A simple analysis shows that this polynomial has at most three or at least one real positive root, depending on the relative values of $H$, $m$ and $q$. In cases where there are two roots, one of them is double, corresponding to an extremal case. 

For the numerical and graphical analysis, it is convenient to define the following dimensionless quantities, 
\begin{equation}
  x =\frac{R}{m},\qquad
  \sigma = \frac{q}{m},\qquad
  h = mH ,
\end{equation}
such that Eq.~\eqref{eq:AH1} for the apparent horizons reads
\begin{equation}\label{eq:AH2}
  h^2 x^4 -x^2 + 2 x - \sigma^2 = 0\,. 
\end{equation}

The number of horizons is determined by the parameters $\left( h, \sigma^2 \right)$, and there are four possibilities, corresponding to four regions in the parameter space, as seen in Fig.\ \ref{fig:sigma}, namely:

\begin{description}
\item[Region I] undercharged cases with the formation of two horizons, $R_-$ and $R_+$.
\item[Region II] also undercharged cases but without horizons and therefore presenting a naked singularity.
\item[Region III] slightly overcharged cases with the formation of three horizons, $R_0$, $R_-$ and $R_+$.
\item[Region IV] overcharged cases with one single horizon $R_+$. 
\end{description}
 
It is important to note that a time-dependent solution, with $H = H(t)$, may correspond to different regions at different times. In the cases we consider, as $\lim_{t \to 0} H = \infty$, $\dot{H} < 0$ and $q =\, \text{constant}$, the evolution of a given solution in the parameter space describes a horizontal straight line, at the corresponding constant $\sigma^2$, starting from infinity and moving to right, towards smaller values of $h$. Depending on $\sigma^2$ and the final value $\lim_{t \to \infty} H(t)$, such solutions may present a behavior corresponding to one or two regions of the $(h, \sigma^2)$ plane.

The existence of AHs is in a sense related to the singularity problem. As seen in the previous section, the spacelike curvature singularity at finite areal radius only appears if $m^2 \geq q^2$. If $m^2 < q^2$, the only curvature singularity appears at $R = 0$ and it is timelike. For this reason, we will analyze the relevant cases separately.

\subsubsection{ Regions I and II}

Regions I and II cover the whole region of the parameter space for $\sigma^2 < 1$, and are separated by the curve for $\sigma_{c-}^2$, the dashed line in Fig.~\ref{fig:sigma}. Region I is a bounded region, while region II is unbounded, extending to $h \to \infty$.

In this case, we may have between one end three real roots for Eq.\ \eqref{eq:AH2}, but the smallest of them is always below the curvature singularity, which, in terms of reduced variables, is the smallest root of
\begin{equation}\label{eq:singularity2}
  x^2 -2 x + \sigma^2 = 0. 
\end{equation}
This means that there are between zero and two apparent horizons in the region covered by the $(t, R)$ coordinates, as we can see in the example from Fig.\ \ref{fig:geod1}. The whole region II presents no AHs and the singularity at $R_+$ is naked. The geodesic lines for a representative case of such a region are shown in Fig.~\ref{fig:geod2}.

\subsubsection{The line $\sigma^2= 1$, $q^2 = m^2$}

Consider now  the special case $m = q$ ($\sigma^2 = 1$). In this case, the AHs correspond to roots of the two second-order polynomials,
\begin{equation}\label{eq:AHx}
  H R^2 + R - m = 0 \, ,\quad
  HR^2 - R + m = 0\,, 
\end{equation}
which can be easily computed. The solutions to $\eqref{eq:AHx}$ are, respectively,
\begin{align} \label{eq:AHxsol}
  R_{1,2} =&\, \frac{1}{H} \left( -1 \pm \sqrt{1 + 4 m H} \right),\\
  R_{3,4} =&\, \frac{1}{H} \left( 1 \pm \sqrt{1 - 4 m H} \right),
\end{align}
showing that, for  $0 < h \leq \frac{1}{4}$, there is just one real positive root for $R$ smaller than the radius of the singularity locus (which is at $R = m$) and two real roots for $R$ larger than $m$. Hence, there may be two (or none) AHs. For $ h > \frac{1}{4}$ there is no AH, since the only positive root is smaller than the singularity radius.

\subsubsection{Regions III and IV}

Region III is a bounded region in the parameter space. It is bounded from below by the line $\sigma^2 = 1$, from above by the curve for $\sigma_{c+}^2$, and from the right by the curve for $\sigma_{c-}^2$. Region IV is an unbounded region in the parameter space. It comprises all the parameter space with $\sigma^2 >1$ not belonging to region III.

When $|q| > m$, Eq.\ \eqref{eq:singularity2} has no roots and there is no finite radius singularity to consider. Therefore, all positive roots of Eq.\ \eqref{eq:AH2} should be considered AHs in the spacetime. Thus, the overcharged case displays three AHs, for solutions in region III (cf. Fig.~\ref{fig:geod3}) or one AH, for solutions in region IV (cf. Fig.~\ref{fig:geod4}). 


%

\subsubsection{Extremal cases}

The boundaries between the various regions in parameter space are given by the extremal cases, where two or more horizons coincide. An analysis of the roots of Eq.~\eqref{eq:AH2} lead us to the graph in Fig.~\ref{fig:sigma}, whose deduction is given in Appendix~\ref{sec:appendix}. The boundaries are given by
\begin{equation}
  \sigma^2_{c\pm} (h) = x_{c\pm}^2 (h) \left[ 1 - 3 h^2 x_{c\pm}^2 (h) \right],
\end{equation}
where
\begin{equation}
  x_{c\pm} (h) =  \frac{\sqrt{6}}{3 h} \cos \left[ \frac{\pi}{3} \pm \frac{1}{3} \arccos \left( \frac{3 \sqrt{6} h}{2} \right) \right].
\end{equation}

\begin{figure}[!htbp]
  \includegraphics[width=0.45\textwidth]{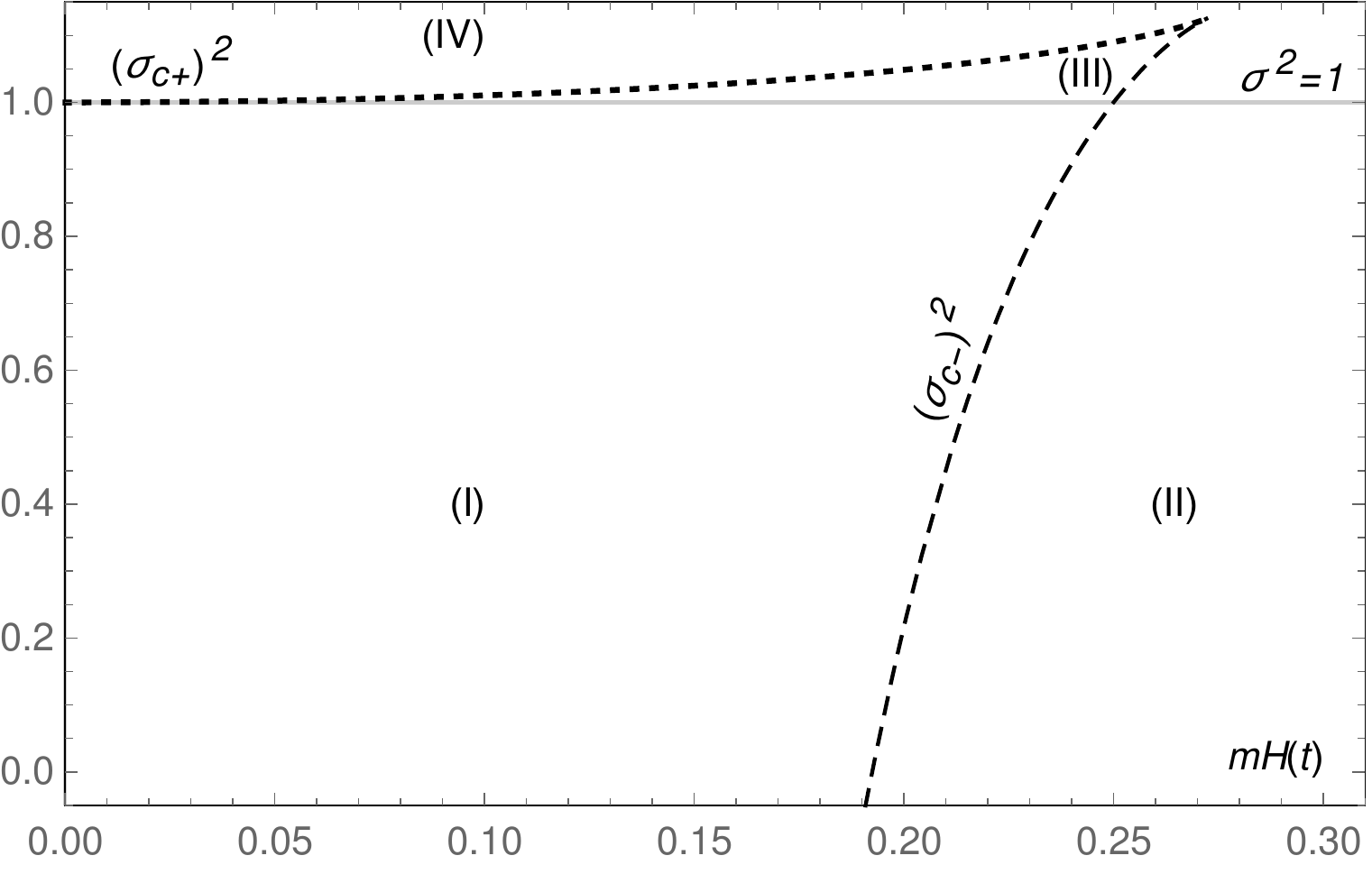}
  \caption{Graphs of the different $\sigma^2 = \frac{q^2}{m^2}$ curves for the extremal cases as a function of $h = m H (t)$: $\sigma_{c+}^2$ (dotted curve) and $\sigma_{c-}^2$ (dashed curve). The solid horizontal line is the $\sigma^2 = 1$ curve, as indicated. Regions I and II cover the whole region of the parameter space for $\sigma^2 < 1$, and are separated from each other by the curve of $\sigma_{c-}^2$.  Region III is bounded by the line $\sigma^2=1$ (from below), and by the curves for $\sigma_{c+}^2$ (from above) and for $\sigma_{c-}^2$ (from the right). Region IV comprises all the parameter space not belonging to the other three regions.} \label{fig:sigma}
\end{figure}

The dotted and dashed curves contain the extremal cases. The dotted line corresponds to the cases where the two innermost AHs coincide, while the dashed line corresponds to cases where the two outermost horizons coincide. The point of convergence of both curves corresponds to the unique case where all three horizons coincide, located at $\sigma^2 = \frac{9}{8}$ and $h = \frac{\sqrt{6}}{9}$.

\subsection{Geodesic completeness}

Let us first consider the case in which, for large enough times, there is a regular region separated from a trapped/anti-trapped region by an AH that we denote $\hat{R}(t)$. Let ${R}_\infty$ be the limiting value of this horizon radius for large times, \emph{i.e.}, ${R}_\infty \equiv \lim_{t \to \infty} \hat{R}_-(t)$. 

Now let us write down the equations for the null ingoing geodesics,
\begin{align}
  \deriv{R_-}{t} =&\, N (R_-) \left[ R_- H (t) - N (R_-) \right],\label{eq:null-geod-t}  \\ 
  R_-'' (\lambda) =&\, \frac{R_- \, R_-'^2 (\lambda) \dot{H} (t)}{N (R_-) \left( N (R_-) - R_- \, H (t) \right)^2},\label{eq:null-geod-lambda}
\end{align}
where the prime indicates derivative with respect to the affine parameter $\lambda$.

\begin{figure}[!htbp]
  \includegraphics[width=0.44\textwidth]{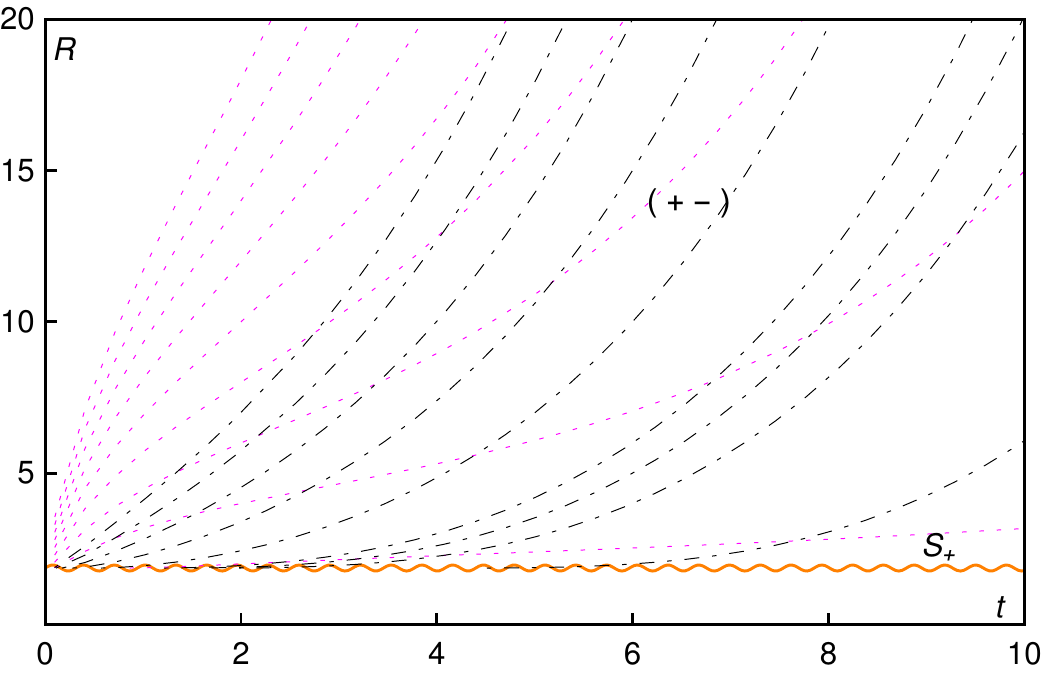}
  \caption{\textbf{Region II:} The ingoing/outgoing geodesics (dotted/dotted-dashed) for the undercharged case when $m = 1$ and $q = 0.50$, considering the scale factor $a (t) = \sinh \left( \frac{3 H_0 t}{2} \right)^{\frac{2}{3}}$, with $H_0 = 0.30$. There is a singularity $S_+$ at $R = m + \sqrt{m^2-q^2}$ but no apparent horizons are formed.}\label{fig:geod2} 
\end{figure}

\begin{figure}[!htbp]
  \includegraphics[width=0.44\textwidth]{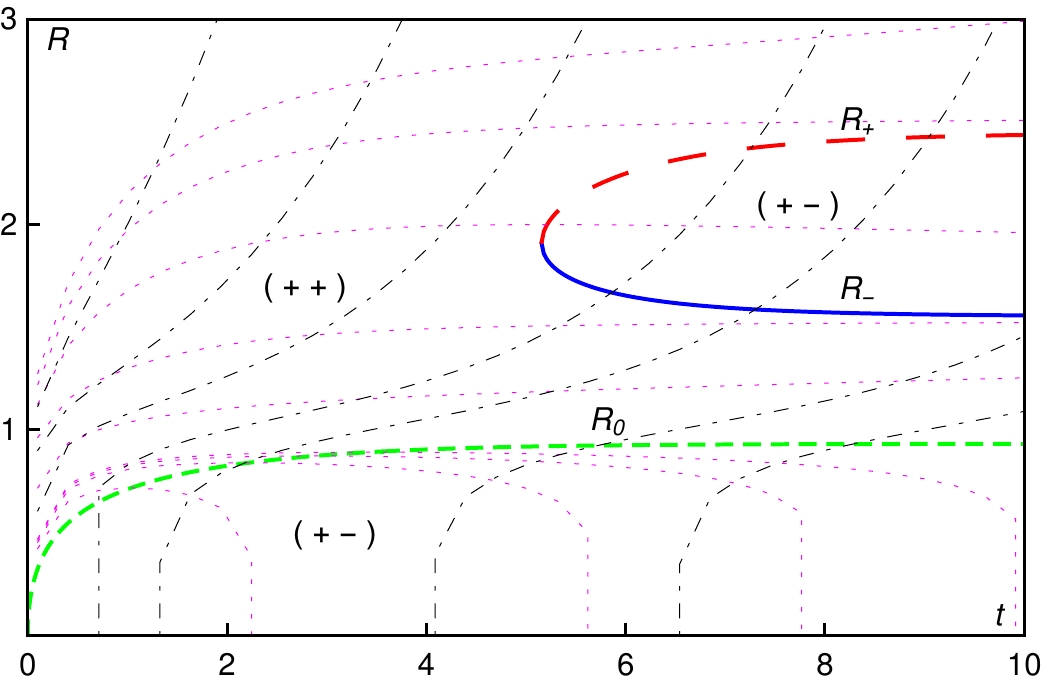}
  \caption{\textbf{Region III:} The ingoing/outgoing geodesics (dotted/dotted-dashed) for the overcharged case when $m = 1$ and $q = 1.02$, considering the scale factor $a (t) = \sinh \left( \frac{3 H_0 t}{2} \right)^{\frac{2}{3}}$, with $H_0 = 0.24$. There are three horizons, indicated by $R_0$, $R_ -$, and $R_+$, that divide the spacetime into two regular regions ($+-$) separated by an anti-trapped region ($++$). There is a singularity at $R=0$, not shown.}\label{fig:geod3} 
\end{figure}

By definition, $\hat{R}(t)$ is a root of $\deriv{R_-}(t)$. If $\deriv[2]{R_-}{t} < 0$, the AH acts as an attractor for ingoing null geodesics that are nearby. In particular, ${R}_\infty = \lim_{t\to \infty} R_-(t)$ for all ingoing null geodesics with values at a neighborhood of ${R}_\infty$. The $\deriv[2]{R_-}{t} < 0$ condition implies that just below the AH ($R < \hat{R}$), $\deriv{R_-}{t} > 0$ and $\deriv{R_-}{t} < 0$ just above the AH ($R > \hat{R}$). This is equivalent to saying that the region below the AH is anti-trapped and the region above it is regular. Such an AH exists in spacetimes that asymptotically reach Regions I and III in Fig.~\ref{fig:sigma}, whose geodesic lines are depicted respectively in Fig.~\ref{fig:geod2} and \ref{fig:geod3}. For those cases, we prove geodesic incompleteness of ingoing null geodesics as follows.

From Eq.\ \eqref{eq:wec}, by imposing the null energy condition over the full spacetime, the Hubble parameter is restricted to $\dot{H} (t) \leq 0$. Therefore, considering a null ingoing geodesic in the regular region, we have $R_-' (0) < 0$ and  $R''_- (0) < 0$, for all $R_- > \hat{R}$. This implies that $R_-$ reaches ${R}_\infty $ at a finite affine parameter interval $\Delta \lambda < \frac{[R_- (0) - {R}_\infty]}{R_-' (0)}$, even though $t \to \infty$. This argument is similar to what was shown in Ref.\ \cite{Kaloper:2010ec} for the uncharged McVittie metric.

When $\deriv[2]{R_-}{t} > 0$, the AH is a repulsive barrier to ingoing null geodesics. This happens for spacetimes that asymptotes to Region III in Fig.~\ref{fig:sigma}. In this case, if there are no more horizons, the ingoing null geodesics below the horizon fall into the central singularity at $R = 0$, being incomplete---for small $R$ the metric behaves as an RNdS spacetime---while the ingoing geodesics above the AH escape to infinity and are complete. When there is no AH the same is valid for ingoing null geodesics, since in this case $\deriv{R_-}{t} > 0$ everywhere, $R_-(t) \to \infty$ as $t \to \infty$, and for large values of $R$ the spacetime behaves as a FLRW spacetime and is therefore geodesically complete at infinity. This is the case for spacetimes that at large times reach Region II of Fig.~\ref{fig:sigma} for large times.

The outgoing null geodesics are always complete and reach future infinity.

\begin{figure}[!htbp]
  \includegraphics[width=0.44\textwidth]{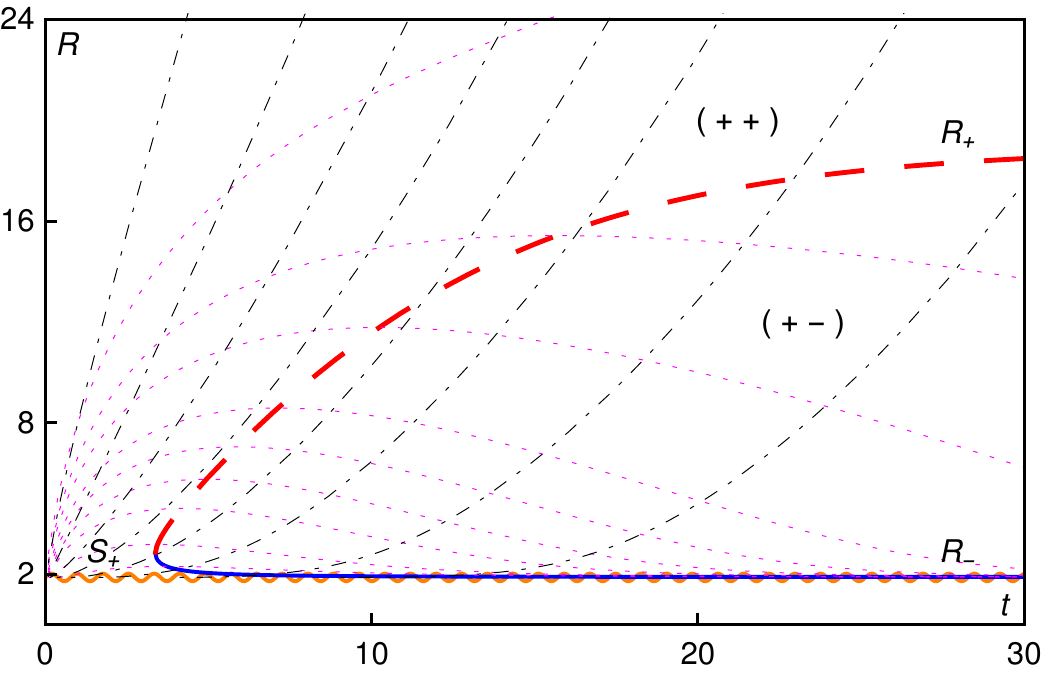}
  \caption{\textbf{Region I:} The ingoing/outgoing geodesics (dotted/dotted-dashed) for the undercharged case when $m = 1$ and $q = 0.50$, considering the scale factor $a (t) = \sinh \left( \frac{3 H_0 t}{2} \right)^{\frac{2}{3}}$, with $H_0 = 0.05$. The singularity in $S_+=m+\sqrt{m^2-q^2}$ is represented by a sinuous line, and the two horizons $R_\pm$ are also represented by continuous and dashed lines, respectively. These horizons divide the spacetime into a regular region ($+-$) and an anti-trapped region ($++$).}\label{fig:geod1} 
\end{figure}

\begin{figure}[!htbp]
  \includegraphics[width=0.44\textwidth]{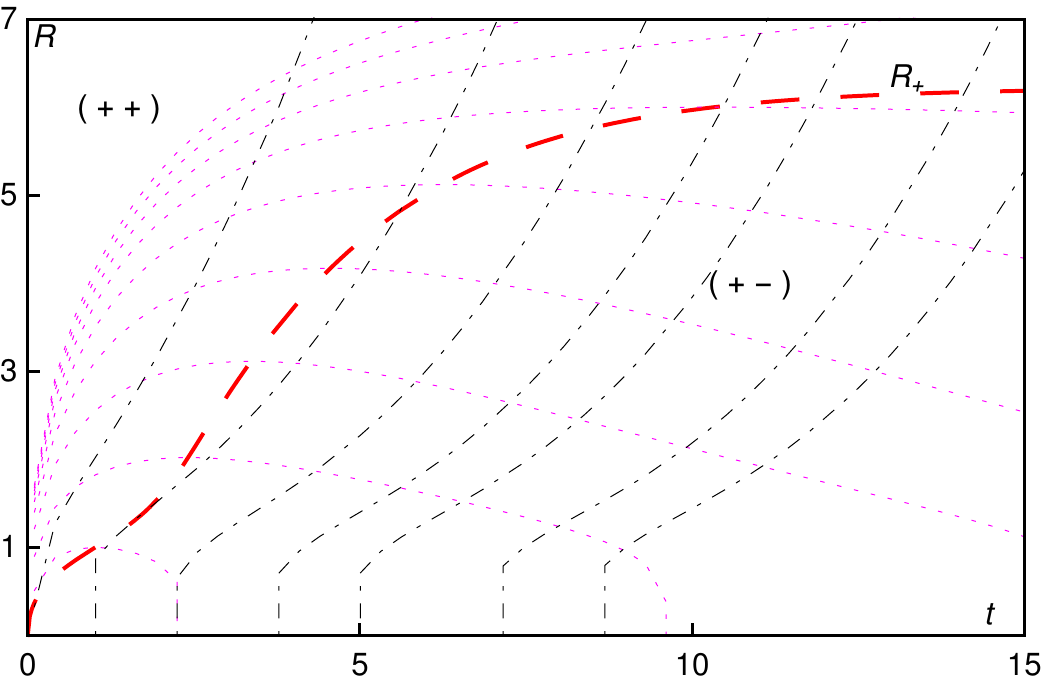}
  \caption{\textbf{Region IV:} The ingoing/outgoing geodesics (dotted/dotted-dashed) for the overcharged case when $m = 1$ and $q = 1.20$, considering the scale factor $a (t) = \sinh \left( \frac{3 H_0 t}{2} \right)^{\frac{2}{3}}$, with $H_0 = 0.14$. There is a singularity at $R=0$, not shown, and the only horizon is represented by a dashed line which divides the spacetime into a regular region ($+-$) and an anti-trapped region ($++$).}\label{fig:geod4} 
\end{figure}

In Figs.~\ref{fig:geod1} to \ref{fig:geod4} we depict examples of spacetimes whose evolution ends in each one of the four regions shown in Fig.~\ref{fig:sigma}.

\subsection{Analytic continuation}\label{sec:analytical}

As seen above, there are instances where the Shah--Vaidya metric is geodesically incomplete, but whose incomplete geodesics do not end in a singularity. This means that those cases may be extendable by the use of new coordinate patches.

If $H(t)$ asymptotes to a constant $H_0 > 0$, the spacetime metric asymptotes to a RNdS metric for $t \to \infty$. Since ingoing null geodesics reach $t  = \infty$ for a finite affine parameter, we can say that at that surface the SV metric coincides with an RNdS metric. In order to prove this, we can extend the SV metric by gluing it with a corresponding RNdS patch at $t = \infty$, for which we need to show that there exists a non singular coordinate system covering the two pieces. This was done for the McVittie solution by Kaloper et al.\ \cite{Kaloper:2010ec} and, following that work, we define the new coordinate
\begin{equation}
  \ud \tau = \ud t + \frac{\ud R}{N(R)\left(N(R) - H_0 R\right)}\,,
\end{equation}
which is formally identical to the construction made in Ref.\ \cite{Kaloper:2010ec}, but for the fact that in our case $N(R)$ contains a charge term. We must prove that the total variation $\Delta \tau$ along ingoing null geodesics close to $\hat{R}_-(t)$ is bounded as $t \to \infty$. Using Eq.~\eqref{eq:null-geod-t} we obtain, along ingoing null geodesics,
\begin{equation}
  \deriv{\tau}{R} =  \frac{1}{N(R)} \left(\frac{1}{N(R) - H_0 R} - \frac{1}{N(R) - H(t) R} \right), 
\end{equation}
which leads to
\begin{equation}\label{eq:deltatau}
  \Delta \tau \!=\!\! \int^{R_\infty}\!\!\!\! \frac{\ud R}{N(R)}\!\! \left( \!\frac{1}{N(R) - H_0 R} \!-\! \frac{1}{N(R) - H(t) R} \!\right).
\end{equation}

The integrand of Eq.~\eqref{eq:deltatau} diverges on the horizon $\hat{R}_-$ and at $R_\infty$. We can make an expansion near $R_\infty$ to obtain, at leading order in $R-R_\infty$,
\begin{equation}\label{eq:tau-first-order}
 \Delta \tau \approx  \int_{z_0}^0 {\ud z} \frac{ - H_0^2 R_\infty^2 \Delta H(t) }{B^2 z^2 - H_0^2 R_\infty^2 B \Delta H(t) z},
\end{equation}
where $z=z(t) \equiv R - R_\infty$, $B \equiv H_0 R_\infty \left[N'(R_\infty) - H_0\right]$, and $\Delta H(t) = H(t) - H_0$. 

Now we must consider the two allegedly small quantities that appear in the integrand of Eq.~\eqref{eq:tau-first-order}, $z$ and $\Delta H$. For $z(t)$ we apply the same expansion procedure to Eq.~\eqref{eq:null-geod-t}, to get $\deriv{z}{t}  \approx \left(H_0 R_\infty^2 \Delta H - B z \right)$, which yields
\begin{equation}\label{eq:z(t)}
z(t) \approx e^{-B(t-t_0)}z_0 + H_0 R_\infty^2 e^{-B t}\!\! \int_{t_0}^{t} e^{B u}\Delta H(u) 
\ud u .
\end{equation}

Following the procedure of Ref.~\cite{daSilva:2015mja}, we consider two possibilities, depending on the asymptotic behavior of $\Delta H(t)$:

\begin{itemize}
\item $\Delta H(t) \in \textit{o} (e^{-B t})$. This means that the first exponential term in Eq.~\eqref{eq:z(t)} dominates for large times, thus $z(t) \approx  e^{-B (t-t_0)}z_0$, and Eq.~\eqref{eq:tau-first-order} becomes
  \begin{equation}\label{eq:delta-tau-hpequeno}
    \Delta \tau \approx \int^0 \frac{\ud z}{z} \frac{\Delta H}{z}.
  \end{equation}
  Therefore, if $\frac{\Delta H}{z} \in \, \textit{o} (z^{\alpha})$, for $\alpha > 0$, then the integral in Eq.~\eqref{eq:delta-tau-hpequeno} converges. This last condition is verified for the most physically meaningful scenarios, which lead to an exponentially vanishing $\Delta H(t)$, that is $\Delta H(t) = \exp(-\beta t)$, with $\beta > B$.

\item $\Delta H(t) > \mathcal{O}(e^{-B t})$. This means that the second term in Eq.~\eqref{eq:z(t)}, the one containing $\Delta H(t)$, dominates the first exponential term for large times. Then, we can approximate $z(t)$ by
  \begin{equation}\label{eq:z-hgrande}
    z(t) \approx \frac{1}{B} \Delta H(t) + e^{-B (t-t_0)} \left(z_0 + \frac{\Delta H(t_0)}{B} \right),
  \end{equation}
  where only the first term dominates for large times. Substituting the leading term of Eq.~\eqref{eq:z-hgrande} into Eq.~\eqref{eq:tau-first-order}, we obtain
  \begin{equation}
    \Delta \tau \sim \int^0 \ud z \frac{1}{\Delta H}.
  \end{equation}
  Since, in this case, $z(t) \sim \Delta H$, we have
  \begin{equation}
    \Delta \tau \sim \int^0  \frac{\ud (\Delta H)}{\Delta H},
  \end{equation}
  which diverges. 

\end{itemize}

Therefore, the coordinate $\tau$ is only suitable in order to analytically extend the Shah--Vaidya spacetime if $e^{-B t}$ dominates $\Delta H(t)$. This situation includes the physically relevant cases of the McVittie spacetime considered in Ref.~\cite{Kaloper:2010ec}, which hold here for the charged case. This does not means that the Shah--Vaidya spacetime is not extendable in other cases, since the ingoing null geodesics are incomplete and there is no singularity where they end the spacetime may be extendable. The meaning of this result is that $\tau$ is not a good coordinate for the extension in those cases, so another coordinate should be employed (a coordinate proportional to the affine parameter of ingoing null geodesics, for example, would be finite).

In the cases where the $\tau$ coordinate is finite as $t \to \infty$, we can extend the spacetime with a patch of RNdS spacetime, which consists in replacing $H(t)$ by $H_0$ in the Shah--Vaidya metric written in the $\tau$ coordinate, namely,
\begin{equation}
  \begin{split}
    \ud s^2 = &- \left(N^2 - H^2 R^2 \right) \ud \tau^2 \\
    -&\frac{2 \ud \tau \ud R}{N} \left( HR - \frac{N^2 - H^2 R^2}{N- H_0 R}\right) \\
    + & \frac{\ud R^2}{N^2}\left( 1 + \frac{2HR}{N - H_0 R} -  \frac{N^2 - H^2 R^2}{\left(N- H_0 R\right)^2} \right) + R^2 \ud \Omega^2.
  \end{split}
\end{equation}

At the limit $R \to R_\infty$ and $ t \to \infty$, the metric reduces to
\begin{equation}
  \ud s^2 = 2 \ud \tau \ud R + R_\infty ^2 \ud \Omega,
\end{equation}
which is the metric of a null surface, as it happens for the uncharged McVittie spacetimes, but now corresponding to one of the event horizons of the static RNdS metric. Which kind of horizon and which patch of the RNdS spacetime appears behind depends on the properties of the RNdS spacetime limit---whether Region I or III in parameter space (cf. Fig. \ref{fig:sigma})---and on the asymptotic properties of the Hubble function $H(t)$ as we discuss next.

\subsection{A causal structure theorem}

In Ref.~\cite{daSilva:2012nh}, a theorem concerned to the causal structure for uncharged McVittie spacetimes was stated and a generalized version of it was proved in Ref.~\cite{daSilva:2015mja}. This theorem specifically tells us that the analytical continuation of the spacetime depends not only on the limit $t \to \infty$, but also on how this limit is approached. The important features to determine in order to analytically continue the spacetime through the null surface at the timelike infinity are the kind of surface it is, and in which direction the ingoing null geodesics are crossing it. In other words, the theorem determines whether the ingoing null geodesics reach a horizon from a regular region or from a trapped/anti-trapped region. Moreover, since the expansion of outgoing null geodesics changes sign at the horizon, and taking into account that the horizon is the boundary separating those regions, the theorem also tells us what kind of region (of the static limiting spacetime) lies behind the horizon.

Here we establish an analogous result available for the Shah--Vaidya metric. We separate the analysis according the region of Fig.~\ref{fig:sigma} each case reaches asymptotically.

\subsubsection{Region I}

Shah--Vaidya spacetimes that, at late times, reach Region I of Fig.~\ref{fig:sigma} are similar to uncharged McVittie spacetimes with respect to singularities (in the patch covered by coordinates $(t,R)$) and AHs, as they have a big-bang singularity at finite areal radius and two AHs, separating a regular region from two anti-trapped regions, as depicted in Fig.~\ref{fig:geod1}. 

For the present study, it is convenient to define the function
\begin{equation}\label{eq:f(t,R)}
  f (t, R) = R H (t) - N (R),
\end{equation}
such that the inner horizon $R_- (t)$ is the smallest solution of $f (t ,R_- (t)) = 0$, that lies in the region above the singular locus $R = R_*$. We also define the quantity 
\begin{equation}
  R_\infty \equiv \lim_{t \to \infty} R_-(t).
\end{equation}
The derivative of the inner horizon function with respect to time $t$ is given by
\begin{equation}
  \dot{R}_-(t) = - \frac{R_-(t) \dot{H}(t)}{f'(t,R_-(t))} < 0.
\end{equation}
The last result is obtained by noticing that $\dot{H} < 0$, $f' (t, R_-) < 0$, following the same line of arguments given in \cite{daSilva:2012nh}. This guarantees that, following the argument of Ref.\ \cite{daSilva:2015mja}, when the areal radius of the inner apparent horizon is decreasing, there may be only two types of causal structure: a single black hole or a black-hole/white-hole pair.

Applying the same method for the Causal Structure Theorem (CST) used in Refs.\ \cite{daSilva:2012nh,daSilva:2015mja}, we obtain the following result for $\lim_{t \to \infty} H (t) = H_0 > 0$:

\begin{proposition} \label{prop}
  Let there be $\Delta H = H - H_0$, $B =- N(R_\infty)f'_\infty (R_\infty)= \frac{m}{R_\infty^2} - \frac{q^2}{R_\infty^3}-H_0^2  R_\infty$, and $t_i > 0$. If there exists $\delta > 0$ such that
  \begin{equation}
    \mathcal{F}^\delta_- (t_i, t) = \int_{t_i}^{t}  e^{(B-\delta)u} \Delta H(u) \diff u,
  \end{equation}
  diverges as $t \to \infty$, then a single black hole is present. Analogously, if there exists $\delta > 0$ such that
  \begin{equation}
    \mathcal{F}^\delta_+ (t_i, t) = \int_{t_i}^{t}  e^{(B+\delta)u} \Delta H(u) \ud u,
  \end{equation}
  converges as $t \to \infty$, then a black-hole/white-hole pair is present.
\end{proposition}

When there is a single black hole, it means that the surface $t \to \infty$ along ingoing null geodesics corresponds to a black-hole event horizon. In the case when there is a black-hole/white-hole pair, part of such a surface corresponds to a black-hole event horizon, but there is a bifurcation point dividing it, and the other part of the surface corresponds to a white-hole horizon. Comparing with the uncharged case, we remark that the charge only appears through its contribution in $R_\infty < R_\infty^{\text{neutral}}$, that changes the value of the parameter $B$. Formally, using our adapted notation, the CST has the exact same form. Representative causal diagrams for these cases are depicted in Figs.~\ref{fig:causal1} and \ref{fig:causal1b}.

\subsubsection{Region III}

In this case, Proposition~\ref{prop} also applies to indicate what kind of region the ingoing null geodesics reach at the end of the coordinate patch near the $\hat{R}_-$ horizon. However, in this case $\hat{R}_-$ is not the innermost horizon, as we have yet another horizon, the $\hat{R}_0$ horizon, between the inner horizon and a timelike singularity at $R = 0$, as depicted in Fig.~\ref{fig:geod3}. Therefore, the resulting causal structure is different and also depends on the behavior of this extra horizon.

Since the spacetimes belonging to this region of the parameter space are pathological, \emph{e.g.}, they do not contain a Cauchy surface, instead of looking for a general analytical result for this case, we prefer just to show some representative causal structure examples, as in Fig.~\ref{fig:causal3} (see Sec.~\ref{sec:examples} for more details).

\subsubsection{Regions II and IV}

In these cases, the causal structure theorem and the Proposition \ref{prop} do not hold. 

In spacetimes represented by Region II there is no AH, and the patch covered by the coordinates is anti-trapped everywhere, therefore there is no doubt about the causal structure. The geodesic lines are depicted in Fig.~\ref{fig:geod2} and a sketch of the corresponding causal diagrams is shown in Fig.~\ref{fig:causal2}.

Spacetimes represented by Region IV present a single AH dividing separating an anti-trapped region that extends until future infinity from a regular region near the singularity at $R=0$. There is no ambiguity about the final destiny of ingoing null geodesics, since all of those that are in the regular region fall into the singularity, while those in the anti-trapped region reach future infinity. This case is depicted in Figs.~\ref{fig:geod4} and \ref{fig:causal4}.

\subsubsection{Application of the CST to a simple model}

Here we consider models with a Hubble function of the form
\begin{equation}\label{eq:H(t)}
  H(t) = H_0 \coth\left(\frac{3 H_0 t}{2} \right),
\end{equation}
where $H_0$ is a positive constant. 

Following the reasoning of Sec.\ V of Ref.\ \cite{daSilva:2012nh}, we have the asymptotic form for $\mathcal{F}^\delta_\pm$,
\begin{equation}
  \mathcal{F}^\delta_\pm \sim \int e^{(B - 3 H_0 \pm \delta)u} \ud u.
\end{equation}
Therefore, convergence will be determined by the sign of $B- 3H_0$. Again, following the procedure of Ref.\ ~\cite{daSilva:2012nh}, we define the parameter
\begin{equation}
  \eta = \frac{B}{3H_0} - 1,
\end{equation}
which is important to determine the type of causal structure in undercharged SV spacetimes, for which $m^2> q^2$. If $\eta > 0$, then the solution corresponds to the single black hole case and the causal structure corresponds to the one depicted in Fig.~\ref{fig:causal1}. If $\eta < 0$, the solution corresponds to a black-hole/white-hole pair and the causal structure is the one depicted in Fig.~\ref{fig:causal1b}.

\begin{figure}[!htbp]
  \includegraphics[width=0.45\textwidth]{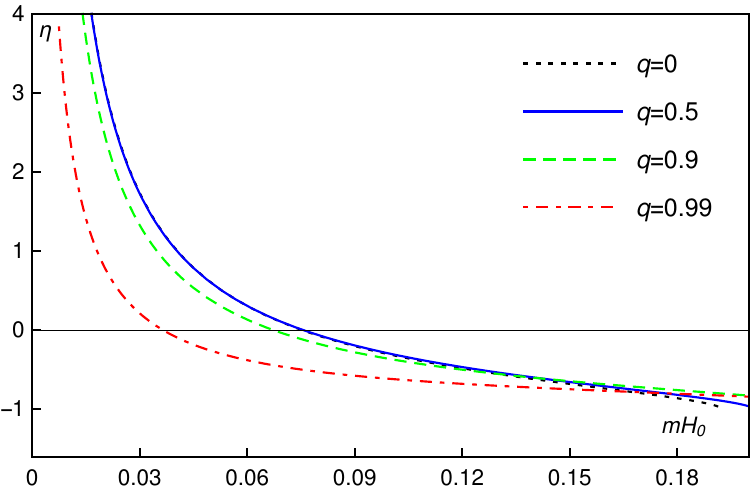} 
  \caption{Behavior of the parameter $\eta$ as a function of $mH_0$ for several values of $q$. We see that the behavior is similar to the uncharged case, with significant changes happening only very close to the extremal case $q = m$. The zeros of the curves for $\eta$ are: $mH_0 = 0.0755$ for $q =0$, $mH_0= 0.0758$ for $q=0.5m$, $mH_0 = 0.0674$ for $q = 0.9m$ and $mH_0 = 0.0366$ for $q = 0.99m$.}\label{fig:eta}
\end{figure}

The value of $\eta$ depends on $h = m H(t)$ and $\sigma = \frac{q}{m}$ for each model, according to Fig.~\ref{fig:eta}. We can see that, even considering only models that behave asymptotically like a $\Lambda$CDM model, we can find both types of causal structures shown in Figs.~\ref{fig:causal1} and~\ref{fig:causal1b} for every value of the charge, provided that $\frac{q}{m} \leq 1$.

\subsection{Causal structure: A few interesting cases}\label{sec:examples}

A few interesting causal structures can be crafted in four different scenarios related with the regions presented in Fig.~\ref{fig:sigma}. As in the previous section, we consider here models whose Hubble parameter is of the form given by Eq.~\eqref{eq:H(t)}.

\subsubsection{Region I}\label{sec:regionI}

This region, whose AHs and geodesic lines are presented in Fig.~\ref{fig:geod1}, represents an undercharged case with $q<m$ and small $mH_0$. For such values of the parameters there is a singularity $\mathcal{S}$ at $R = m + \sqrt{m^2 - q^2}$ that can be considered as an initial singularity, since it is spacelike and no information from its past can escape to the future. This means that, for spacetimes that asymptotically reach Region I, we can use a numerical method explained in Ref.~\cite{walker-1970,Lake:2011ni,daSilva:2012nh} in order to draw the respective causal diagrams. This method consists of the defining a Cauchy surface from which we can connect each event of the spacetime patch under study by an ingoing null geodesic and an outgoing null geodesic. We numerically integrate the two geodesics and perform a coordinate transformation in order to depict a compact space where the null geodesics are straight lines at \SI{45}{\degree}. Plotting the loci of interest in this conformally compact space we obtain the causal structure. In these cases, we define a Cauchy surface slightly above the initial singularity in order to apply the numerical method mentioned. 

\begin{figure}[!htbp]
  \includegraphics[width=0.48\textwidth]{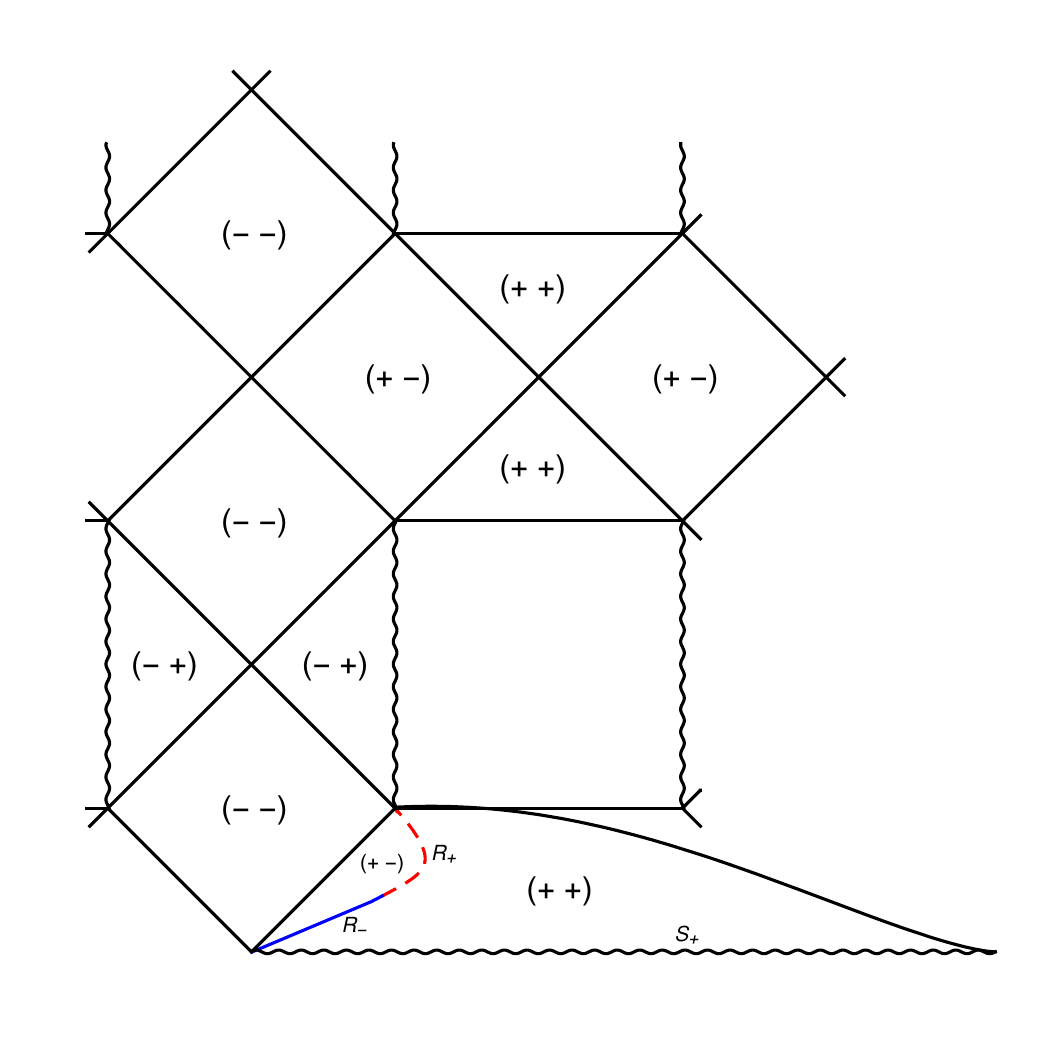}
  \caption{Causal structure for the undercharged SV spacetime with $m=1$, $q=0.5$ and $H_0=0.05$, where there are two apparent horizons covering the singularity at $N(R)=0$ for observers in the regular region. The horizons were found numerically while the boundaries are schematic.}
  \label{fig:causal1}
\end{figure}

The causal diagram presented in Fig.~\ref{fig:causal1} corresponds to the single black hole case of Proposition~\ref{prop}. It presents two AHs, $\hat{R}_-$ and $\hat{R}_+$ respectively, along with an initial singularity at $\mathcal{S}_+$. The null surface at the left correspond to the surface reached at time infinity by ingoing null geodesics. At this surface, the metric is equal to its RNdS limit, meaning that the causal structure can be extended and ``glued'' in a RNdS manifold such as the one presented in \cite{Weiss}, considering that the region behind the event horizon must be a RNdS trapped region, as both null ingoing and outgoing geodesics have negative expansion scalars on that side of the horizon.

\begin{figure}[!htbp]
  \includegraphics[width=0.48\textwidth]{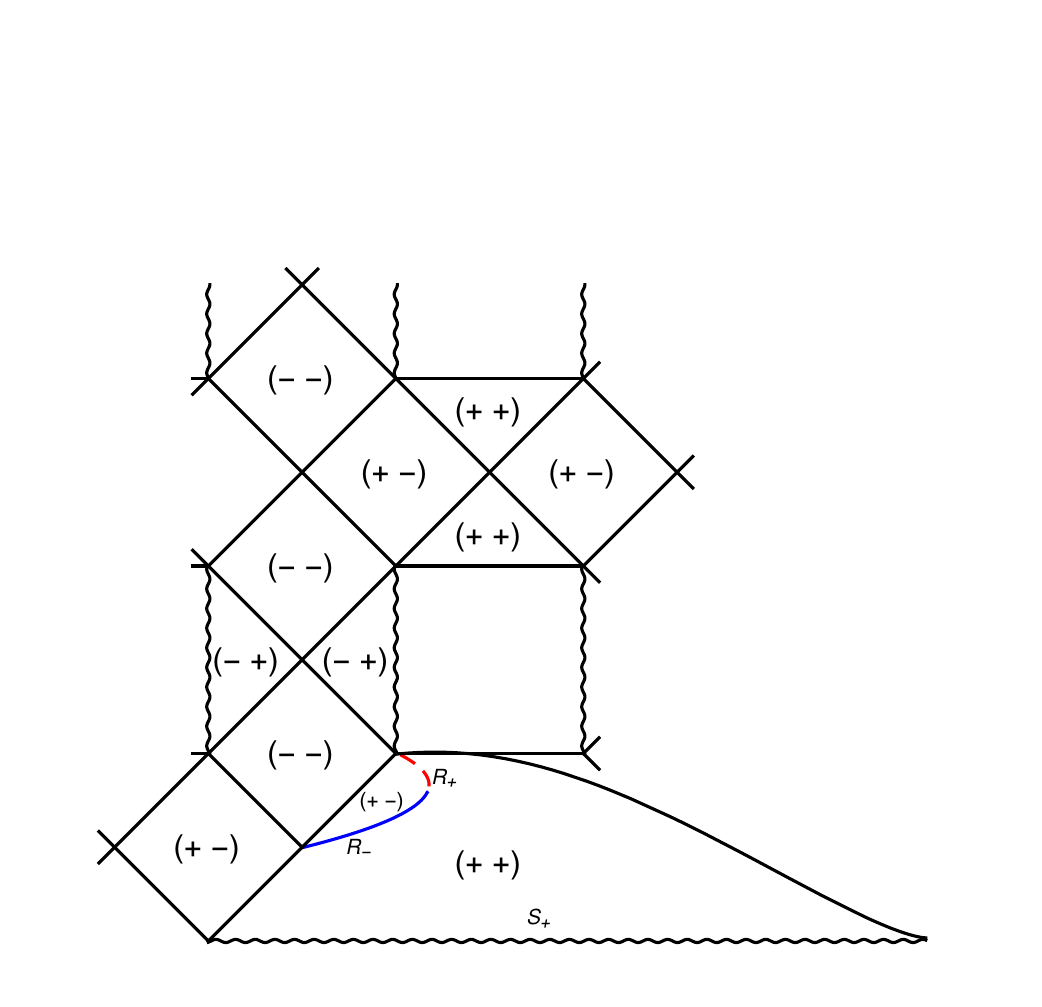}
  \caption{Causal structure for the undercharged SV spacetime where $mH_0>0.075$. There are two apparent horizons and a bifurcation point, besides a singularity at $N(R)=0$ that is covered by the two horizons.}
  \label{fig:causal1b}
\end{figure}

In Fig.~\ref{fig:causal1b}, the continuation is also into a RNdS manifold, but the difference is that part of the null surface at the timelike infinity is not a black hole event horizon, as there are a set of ingoing null geodesics that are able to cross it coming from an anti-trapped region. This means that a regular RNdS patch lies behind this part of the horizon, where the two null congruences have expansions of opposite signs. This surface is a white-hole horizon, whose interior is the anti-trapped region near the singularity in the $(t,R)$ coordinate patch. There is also an event horizon, as the boundary of the regular region, that should be continued into a trapped RNdS region, as in the previous case of Fig.~\ref{fig:causal1}.

\subsubsection{Region II}
\label{sec:regionII}

\begin{figure}[!htbp]
  \includegraphics[width=0.48\textwidth]{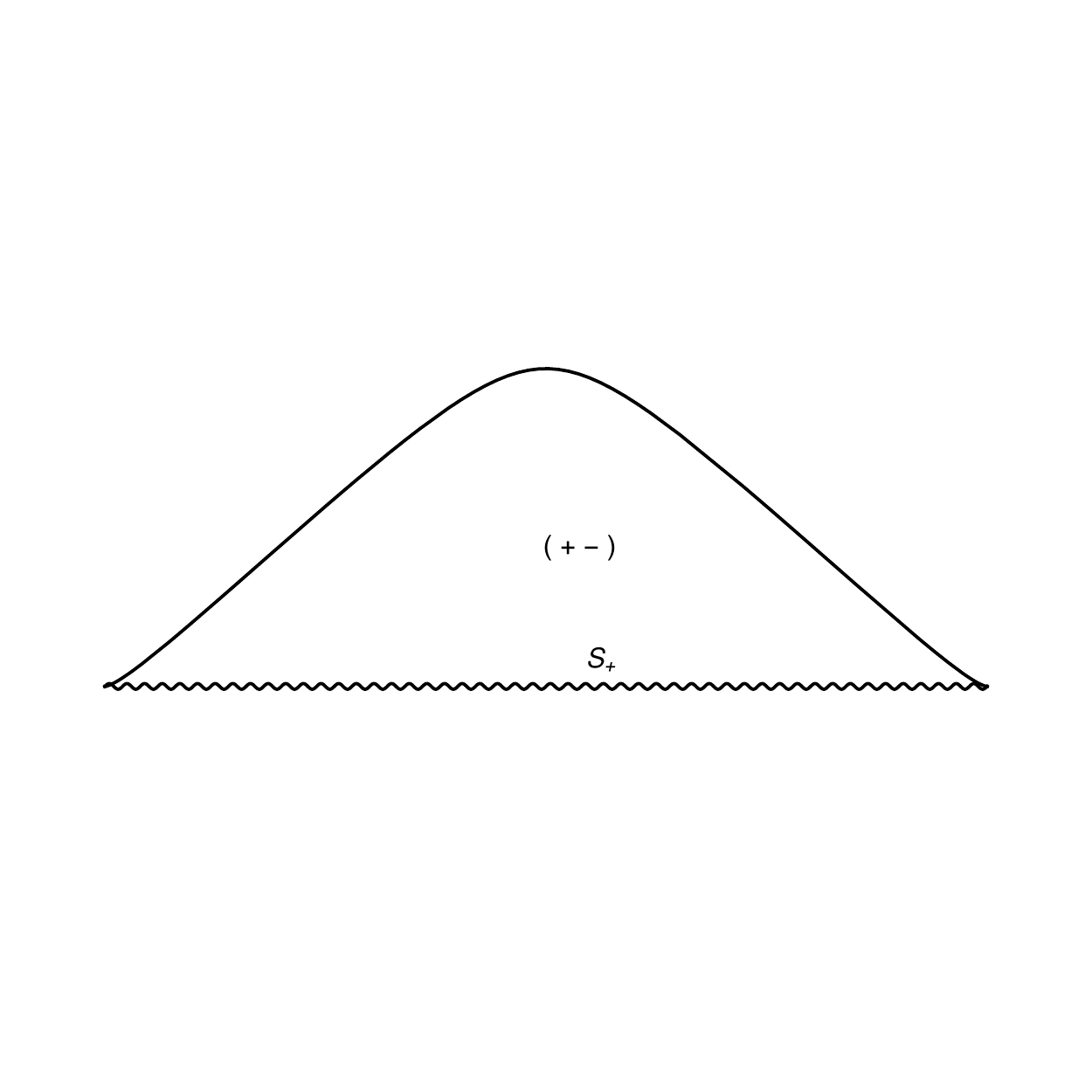}
  \caption{Causal structure for the undercharged SV spacetime without horizons and a naked singularity at $N(R)=0$.}
  \label{fig:causal2}
\end{figure}

The solutions in this region correspond to asymptotically undercharged RNdS spacetimes but now with higher values of $mH_0$. The singularity at $R=m+\sqrt{m^2-q^2}$ is still present, however no AH is formed and the singularity is naked, as can be seen in Fig.~\ref{fig:geod2} and in the causal diagram of Fig.~\ref{fig:causal2}. These spacetimes are not analytically extendable. 
 
\subsubsection{Region III}

\begin{figure}[!htbp]
  \includegraphics[width=0.48\textwidth]{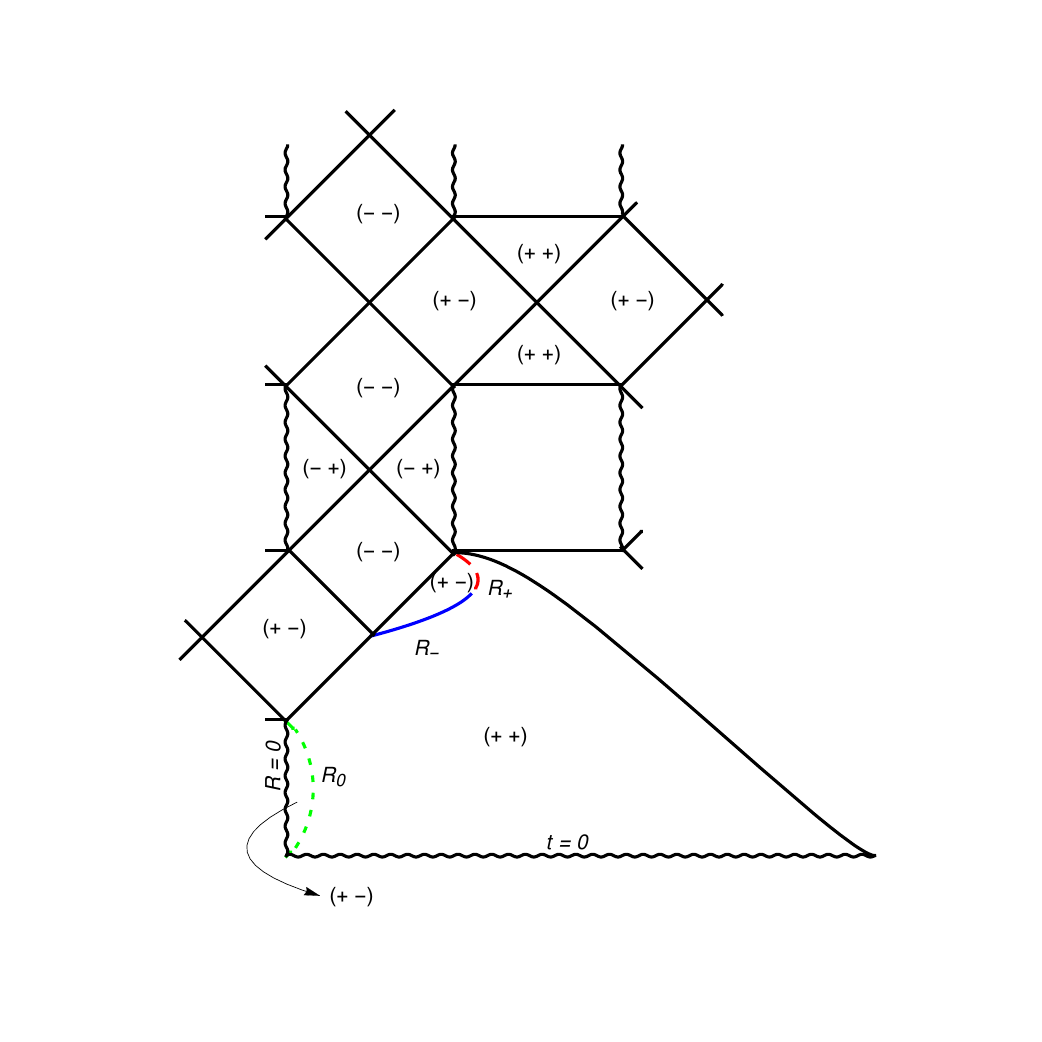}
  \caption{Causal structure for the overcharged SV spacetime with three horizons and a singularity at $R = 0$.} 
  \label{fig:causal3}
\end{figure}

The solutions in this region correspond to spacetimes that asymptote to overcharged RNdS solution with $1 < \frac{q}{m} < \num{1.125}$. They present three AHs, a singularity at $R = 0$ and an initial singularity at $t = 0$, as shown in Fig.~\ref{fig:geod3}. These cases do not present a Cauchy surface and a numerical construction of the horizons is not possible, therefore the whole diagram in Fig.~\ref{fig:causal3} is schematic. The inner horizon $R_0$ covers the singularity in $R = 0$, while the two other horizons, a black-hole one and a cosmological one ($R_-$ and $R_+$ respectively), delimit a regular region that ends in an event horizon, allowing an extension to a RNdS \cite{Weiss} causal structure. The singularity at $R = 0$ is naked, since it is causally connected to external observers. This kind of structure occurs solely with the addition of charge, and is not present in the original McVittie solution.

\subsubsection{Region IV}

\begin{figure}[!htbp]
  \includegraphics[width=0.48\textwidth]{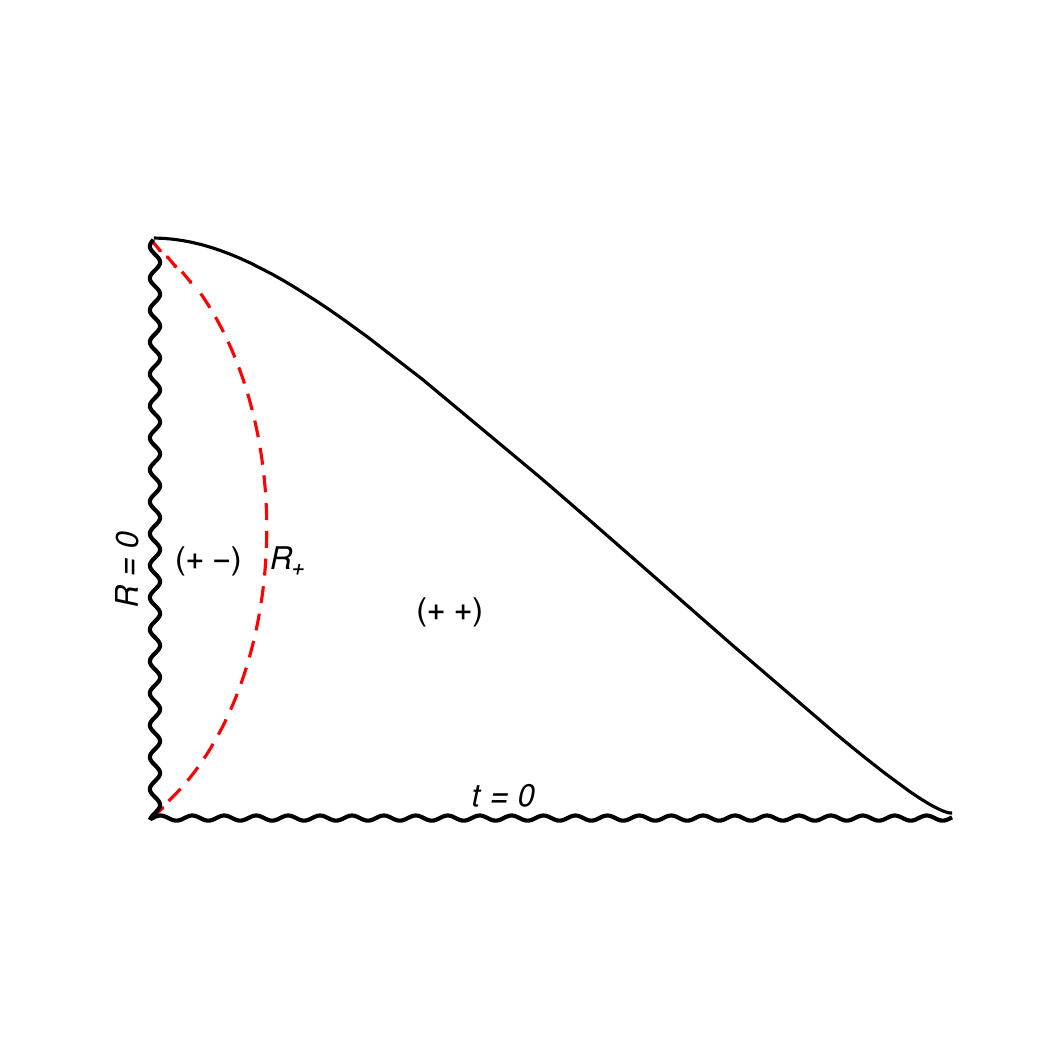}
  \caption{Causal structure for the overcharged SV spacetime with one horizon and a singularity at $R = 0$.} 
  \label{fig:causal4}
\end{figure}

Finally, the fourth region contains overcharged SV spacetimes which also have a singularity at $R = 0$ and an initial singularity at $t = 0$. The Cauchy surface is not present either, as it can be seen in Fig.~\ref{fig:geod4}. These solutions present one cosmological horizon ($R_-$) that covers the singularity at $R=0$ and is causally complete, as showed in the schematic diagram depicted in Fig.~\ref{fig:causal4}.   

\section{Conclusions}\label{sec:conclusion}

In this work, we have studied the Shah--Vaidya solution, which can be understood as a charged version of the known McVittie solution, but considering the presence of a central electric charge. 

For the first time this metric has been derived from a Lagrangian formalism, with the action defined by the Einstein--Hilbert action plus a Maxwell field and cuscuton field as sources. We considered metrics that asymptote to the Reissner--Nordström--de~Sitter solution as $t \to \infty$. We studied the properties of the spacetimes and analyzed the changes of such properties in the parameter space of the asymptotic metric, corresponding to specific values of the electric charge and of the Hubble parameter. We found out that there are regions in the parameter space where the Shah--Vaidya metric does correspond to charged cosmological black holes, provided that the parameters asymptotes to a suitable region of values. On the other hand, some cases correspond to naked singularities.

Regarding the black-hole cases, we concluded that those include the two kinds of structures found in the uncharged McVittie case, with the only difference being that they are analytically continued with patches of Reissner--Nordström--de~Sitter spacetimes, instead of the Kottler spacetimes of the uncharged case. The difference between these two cases does not lie in the limiting values of the metric, but in the asymptotic behavior of the Hubble factor $H(t)$. This is similar to what happens in the uncharged McVittie case, for which a theorem proved in Ref.\ \cite{daSilva:2012nh} determines the asymptotic conditions for each type of causal structure. We have stated here a very similar theorem valid for a large class of the Shah--Vaidya spacetimes. Finally, we provided examples representing each qualitative type of causal structure that can be found under our initial assumptions and constructed the corresponding causal diagrams.

\begin{acknowledgments}

  We thank N.\ Afshordi, L.\ Smolin, R.\ Sorkin and S.\ Surya for insights and valuable discussions. This research was supported in part by Perimeter Institute for Theoretical Physics. Research at Perimeter Institute is supported by the Government of Canada through Innovation, Science and Economic Development Canada and by the Province of Ontario through the Ministry of Research, Innovation and Science. V.\ T.\ Z.\ thanks  \foreignlanguage{brazil}{Coordenação de Aperfeiçoamento de Pessoal de Nível Superior} (CAPES), Brazil, Grants No.~88881.064999/2014-01 and No.~88881.310352/2018-01, and \foreignlanguage{brazil}{Conselho Nacional de Desenvolvimento Científico e Tecnológico} (CNPq), Brazil, Grants No.~308346/2015-7 and No.~309609/2018-6. A.\ M.\ was partially supported by CNPq Grant No.~400342/2017-0. This study was financed in part by CAPES Finance Code 001.
  
\end{acknowledgments}

\appendix

\section{Regions of the parameter space} \label{sec:appendix}

In this Appendix, we study the cases in which the horizons are multiple roots of $P (x)$ defined in Eq.~\eqref{eq:AH2},
\begin{equation}
  P(x) = -h^2 x^4 + x^2 - 2 x + \sigma^2.
\end{equation}
We are interested in the extremal cases, in which two or more horizons coincide; that is, $P (x)$ has a double or triple root. An extremal case is the division between regimes with different properties and possibly a different numbers of horizons. For this purpose, we adopt the following procedure.

\begin{enumerate}[(1)]
\item Fix a value for $h$;
\item Solve the equation $P' (x) = 0$ and obtain the roots $x_{c\pm}$;
\item Solve the equation $P (x_{c\pm}) = 0$ for $\sigma$ in order to find the values $\sigma_ {c\pm}$ that correspond the extremal cases of double roots. 
\end{enumerate}

Therefore, for a double root case, the system we have to solve is the following
\begin{align}
  P (x) =&\, -h^2 x^4 + x^2 - 2 x + \sigma^2 = 0,  \\
  P' (x) =&\, -4 h^2 x^3 + 2 x - 2 = 0. \label{eq:SdS}
\end{align}

We notice that Eq.\ \eqref{eq:SdS} is independent of the charge parameter $\sigma$ and can be solved for $x_c$, the value(s) for which the graphic of $P (x)$ has horizontal slope curve. The solution may be obtained by comparison to the equation for the horizons of Schwarzschild--de~Sitter spacetimes, which yield the same equation as \eqref{eq:SdS}. Therefore, the solution for the roots of $P' (x)$, for $0 < h <  \frac{\sqrt{6}}{9}$ may be written in the form
\begin{equation}
  x_{c\pm}= \frac{\sqrt{6}}{3 h} \cos \left( \frac{\pi \pm \xi}{3} \right),
\end{equation}
where
\begin{equation}\label{eq:xi}
  \xi = \arccos \left( \frac{3\sqrt{6} h}{2} \right).
\end{equation}

We note here that, for $h = \frac{\sqrt{6}}{9}$, the polynomial $P'(x)$ has a double root and, therefore, if $P(x)$ has a root at the same point, it will be a triple root. 

Let $\sigma_{c\pm}$ be a critical charge for which there is a double root at $x_{c\pm}$. Thus, $\sigma_{c\pm}$ satisfies
\begin{equation}
  \begin{split}
    P(x_{c\pm}) = 0 \Rightarrow \sigma_{c\pm}^2 =&\, h^2 x_{c\pm}^4 - x_{c\pm}^2 + 2 x_{c\pm} \\
    =&\, x_{c\pm}^2 \left( 1 - 3 h^2 x_{c\pm}^2 \right),
  \end{split}
\end{equation}
where we used the fact that $P' (x_{c\pm}) = 0$ in order to find the last equality. 

Now, we have to study the range of $h$ for which a double root may exist. The first constraint is that $\sigma_{c\pm}^2 \geq 0$, which implies in $3 h^2 x_{c\pm}^2 \leq 1$. When considering the minus sign, we obtain $h \geq \frac{1}{3\sqrt{3}}$, while for the plus sign we get $|h| < \frac{\sqrt{2}}{3 \sqrt{3}}$. 

Lastly, we should remark that we can build extremal horizons in two ways in this spacetime, as follows.

\begin{enumerate}[(1)]
\item When the internal horizon and event horizon coincide, which corresponds to $x_{c+}$ being a double root of $P (x)$. The range of existence of such a solution is, by combining the constraints found, $0 \leq 3 \sqrt{3}\, h \leq \sqrt{2}$. 
\item When the event horizon and the cosmological horizon coincide, which corresponds to $x_{c-}$ being a double root of $P (x)$. The corresponding range of existence is $1 \leq 3 \sqrt{3}\, h \leq \sqrt{2}$. 
\end{enumerate}

\section{The sources of the Shah--Vaidya spacetime}

For the sake of completeness, we write here the nontrivial components of the Einstein tensor in terms of the areal radius,  
\begin{align}
  \tensor{G}{^{t}_{t}} =& -3 H^2 - \frac{q^2}{R^4},\\
  \tensor{G}{^{R}_{t}} =&\, 2 R H \dot{H},\\
  \tensor{G}{^{R}_{R}} =& -3 H^2 - \frac{q^2}{R^4} - \frac{2 \dot{H}}{N},\\
  \tensor{G}{^{\varphi}_{\varphi}} =&\, \tensor{G}{^{\theta}_{\theta}} = -3 H^2 + \frac{q^2}{R^4} - \frac{2 \dot{H}}{N}.
\end{align}
Therefore, the sources for the Shah--Vaidya metric \eqref{eq:metric-areal} may be interpreted as a perfect fluid whose energy density $\rho$ and pressure $p$ are given respectively by
\begin{align}
  \rho =&\, \frac{3 H^2}{8\pi}+\frac{q^2}{8\pi R^4}  \label{eq:edensity}\\
  p =& -\frac{3 H^2}{8\pi} -\frac{ \dot{H}}{4\pi N}+\frac{q^2}{24\pi R^4}. \label{pressure}
\end{align}
In addition, there is an electromagnetic energy-momentum tensor $E_\mu^\nu$ given by
\begin{equation}
  E_\mu^\nu = \frac{q^2}{R^4} \, \diag \left( -1, 1, -1, -1 \right).
\end{equation}

When establishing the structural properties of the spacetime, such as the presence of singularities and the nature of trapping horizons, an important issue is the analysis of the energy conditions. Useful here are the null (NEC), weak (WEC) and strong (SEC) energy conditions. For the present case we have 
\begin{align}
  \text{NEC:}& \;\; -\frac{\dot{H}}{4\pi N}+\frac{q^2}{6\pi R^4} \geq 0, \label{eq:nec}\\
  \text{WEC:}& \;\; \frac{3H^2}{8\pi} + \frac{q^2}{8\pi R^4} \geq 0, \quad -\frac{\dot{H}}{4\pi N}+\frac{q^2}{6\pi R^4} \geq 0,  \label{eq:wec}\\
  \begin{split}
    \text{SEC:}& \;\; -\frac{3H^2}{4\pi}-\frac{3\dot{H}}{4\pi N} + \frac{q^2}{6\pi R^4} \geq 0, \\
    &-\frac{\dot{H}}{4\pi N}+\frac{q^2}{6\pi R^4} \geq 0. \label{eq:sec}
  \end{split}
\end{align}

Relevant studies were performed in Refs.~\cite{Hayward:1993mw,Andersson:2005gq}, which present theorems relating the chronological character of the apparent/trapping horizons to the null energy condition. Here, for the sake of simplicity, we stick to the statement given in Ref.~\cite{Hayward:1993mw}:

\begin{theorem}\label{theorem}
If the null energy condition holds, then an outer (inner) trapping horizon is spacelike (timelike), and a trapping horizon is null if and only if, additionally, the internal shear and normal energy density (energy flux) vanish.
\end{theorem}

This theorem gives a different way of establishing---indirectly---the character of the apparent horizons in the Shah--Vaidya spacetime.

\section{The case $\lim_{t\to\infty} H(t)=0$}\label{appendixH0zero}

\subsection{Singularities and horizons}

Assuming that the scale factor is a power law function of the form $a(t) = a_0 + a_1 t^\alpha$, with $\alpha > 0$, it follows that $\lim_{t\to\infty} H(t)=0$ and $\lim_{t\to\infty} \dot H(t)=0$. This case corresponds to the vertical axis in Fig.~\ref{fig:sigma}, and the SV metric \eqref{eq:metric-areal} asymptotes to the Reissner--Nordström metric at large times.

From relations \eqref{eq:ricci} and \eqref{eq:kretsch} one concludes that $\displaystyle{\lim_{t\to\infty} {\cal R}(t) =0}$ and $\displaystyle{\lim_{t\to\infty} {\cal K}(t) = 48 \left( \frac{m}{R^3} - \frac{q^2}{R^4} \right)^2 + \frac{8q^4}{R^8}} $. These results confirm that the spacetime presents the same singularities as the Reissner--Nordström solution, a timelike singularity at $R=0$ alone. Since the asymptotic limit is also the Reissner--Nordström solution, a similar causal structure is also expected.

\subsection{Causal structure}

The causal structure presents three qualitatively disjoint cases, all of them asymptotic to the Reissner--Nordström solutions.

\begin{figure}[!htbp]
  \centering
  \includegraphics[width=0.48\textwidth]{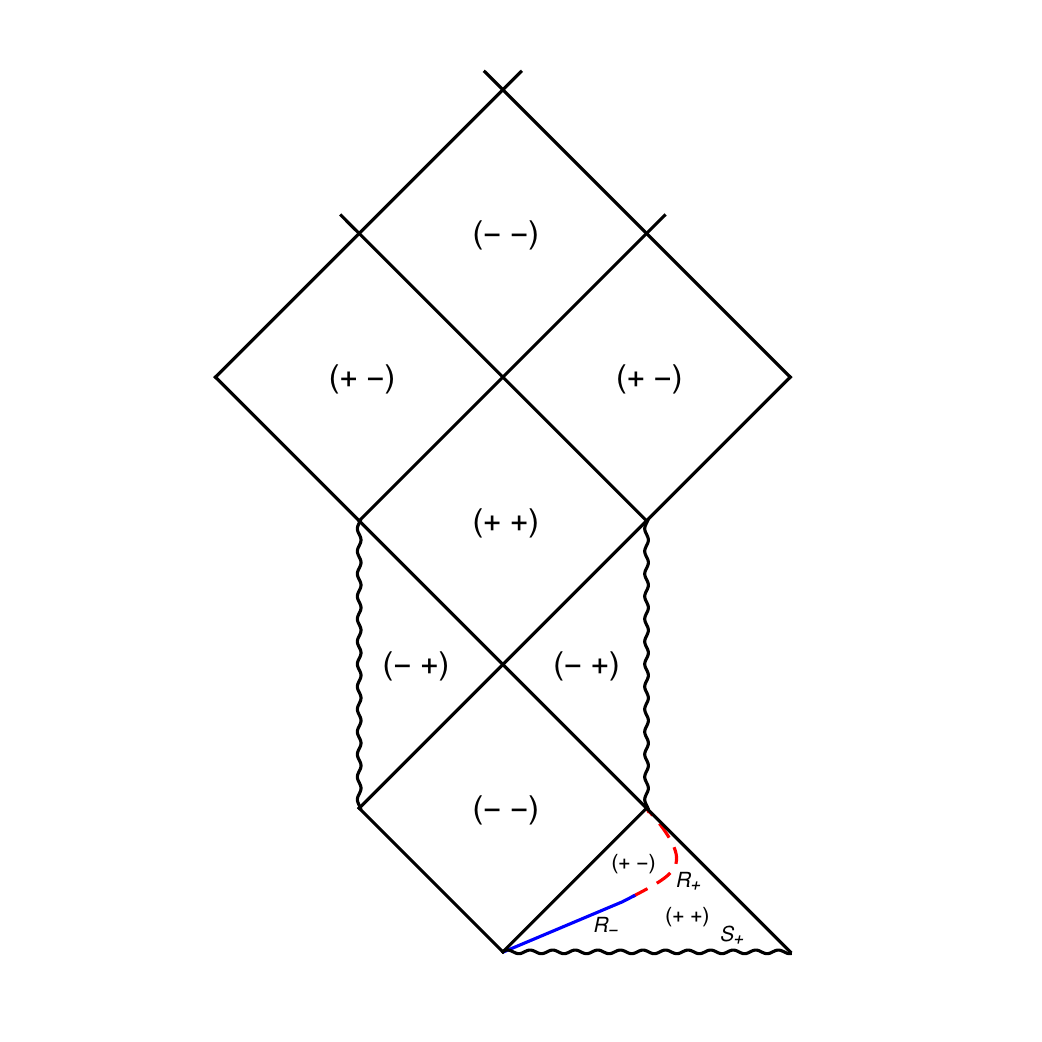}
  \caption{Causal structure for SV spacetime in the case  $a(t) = t^{2/3}$, and with $m^2>q^2$.} 
  \label{fig:causal5}
  \end{figure}

The first case, for values of $m^2>q^2$, presents two apparent horizons $R_+$ and $R_-$ and an initial singularity at $S_+$. The corresponding causal diagram is shown in Fig.~\ref{fig:causal5}. The initial singularity at $t=0$ is hidden to observers in the regular region $(+,-)$ by the two apparent horizons, while observers in the anti-trapped region $(+,\,+)$ are not.

\begin{figure}[!htbp]
  \centering
  \includegraphics[width=0.3\textwidth]{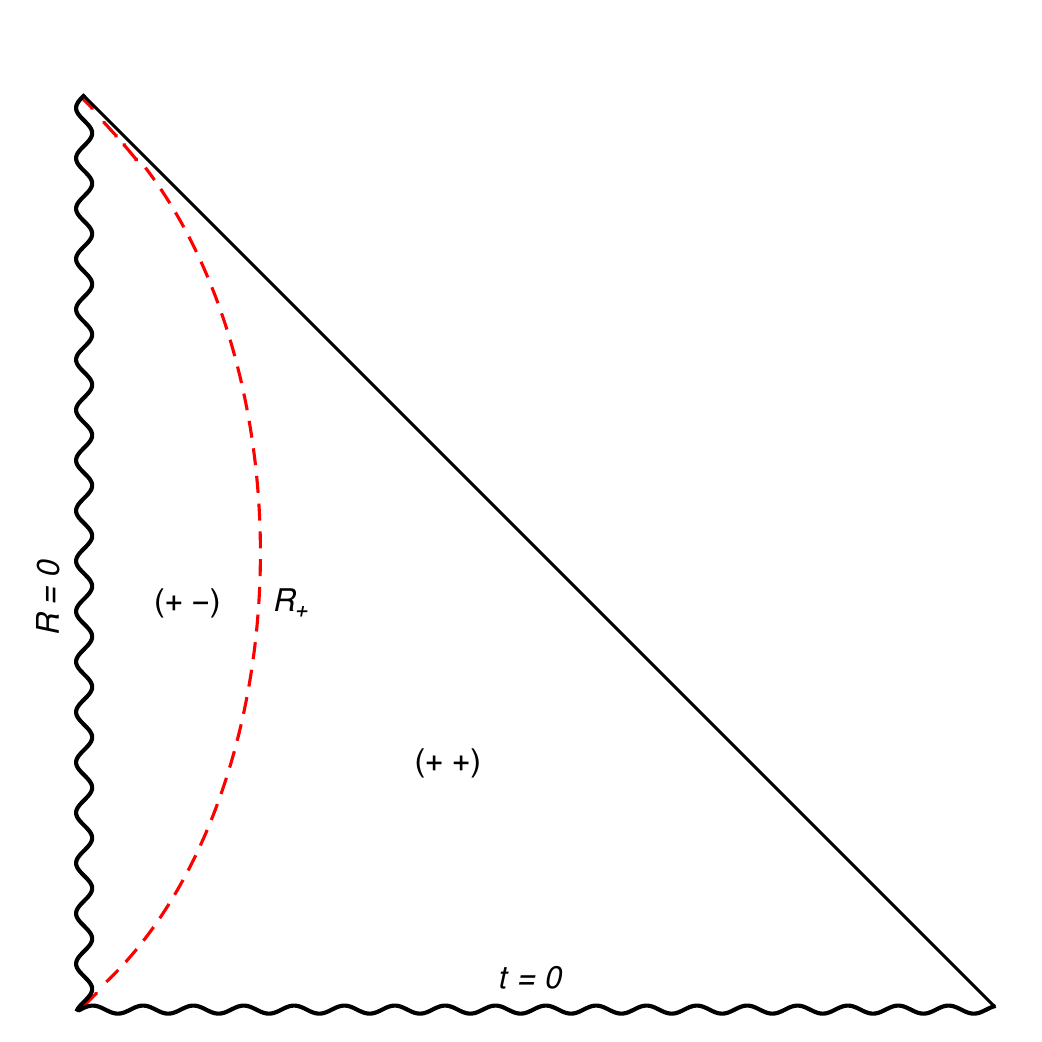}
  \caption{Causal structure for SV spacetime in the case $a(t) = t^{2/3}$, and with $m^2<q^2$.} 
  \label{fig:causal6}
  \end{figure}

The second case is for $m^2<q^2$ and sufficiently far from the line $\sigma =q^2/m^2 =1$, the overcharged case, with one apparent horizon $R_-$ covering the singularity at $R = 0$. Notice that, unlike the $\Lambda$CDM cases, this case presents a lightlike future infinity. The corresponding causal diagram is shown in Fig.~\ref{fig:causal6}. The singularity is hidden to observers in the regular region $(+,-)$ by the two apparent horizons. 

The third case emerges when considering values $q/m \gtrsim 1$, a case with three apparent horizons at intermediate times. Given that this is a very peculiar case that, asymptotically, tends to the case with only one horizon shown in Fig.~\ref{fig:causal6}, we chose to not explore it further.

\bibliographystyle{apsrev4-1}

\bibliography{shortnames,referencias2}

\end{document}